\documentclass[
 nopreprint,
 author-year,
]{jasatex}

\usepackage{graphicx}
\usepackage{dcolumn}
\usepackage{amsmath,amsfonts}
\usepackage{bm}

\usepackage{amssymb,url}
\usepackage{float}
\usepackage{makeidx}
\usepackage{pstricks}
\usepackage[T1]{fontenc}
\usepackage[caption=false]{subfig}
\usepackage{multirow}

\def\c{\noindent}

\def\({\left(}
\def\){\right)}

\def\[{\left[}
\def\]{\right]}

\def\phib{\bm \phi}

\def\n{{\bm n}}

\def\x{{\bm x}}

\def\P{\bm P}

\def\Phii{\bm \Phi_i}
\def\Phiidot {\dot{\bm\Phi}_{i}}
\def\Psib{\bm \Psi}

\def\n{\bm n}

\def\M{\bm M}
\def\Ma{\bm M_{\alpha}}
\def\Mb{\bm M_{\beta}}
\def\Mg{\bm M_{\gamma}}

\def\Mxii{\bm M_{\xi_i}}
\def\K{\bm K}
\def\L{\bm L}
\def\B{\bm B}

\def\Bi{\bm B_i}

\def\Biai{\bm B_{i,a_i}}
\def\Bibi{\bm B_{i,b_i}}

\def\GammaG{\Gamma_{\rm G}}        
\def\GammaW{\Gamma_{\rm W}}        
\def\GammaZ{\Gamma_{\rm Z}}        
\def\GammaH{\Gamma_{\rm H}}        
\def\GammaI{\Gamma_{\rm \infty}}   

\def\Omegae{\Omega_e}

\def\Acta{Acta Acust.}

\def\Ast{Acoust. Sci. \& Tech.}

\def\Cmame{Comput. Methods Appl. Mech. Engrg.}

\def\Jasa{J. Acoust. Soc. Am.}

\def\Jcp{J. Comput. Phys.}

\def\Jsv{J. Sound Vib.}

\def\Mssp{Mech. Syst. Signal Pr.}

\begin{document}

\preprint{AIP/123-QED}

\title[M. Arnela and O. Guasch.: Elliptical vocal tract impedance computation]{Finite element computation of elliptical vocal tract impedances using the two-microphone transfer function method}

\author{Marc Arnela}

\author{Oriol Guasch}%
 \email{oguasch@salle.url.edu}

\affiliation{GTM Grup de recerca en Tecnologies M\`edia, La Salle, Universitat Ramon Llull, C/Quatre Camins 2, Barcelona 08022, Catalonia, Spain\\}

\date{\today}


\begin{abstract}
\c
The experimental two-microphone transfer function method (TMTF) is adapted to the numerical framework to compute the radiation and input impedances of three-dimensional vocal tracts of elliptical cross section. In its simplest version, the TMTF method only requires measuring the acoustic pressure at two points in an impedance duct and the postprocessing of the corresponding transfer function. However, some considerations are to be taken into account when using the TMTF method in the numerical context, which constitute the main objective of this paper. In particular, the importance of including absorption at the impedance duct walls to avoid lengthy numerical simulations is discussed and analytical complex axial wave numbers for elliptical ducts are derived for this purpose. It is also shown how the plane wave restriction of the TMTF method can be circumvented to some extent by appropriate location of the virtual microphones, thus extending the method frequency range of validity. Virtual microphone spacing is also discussed on the basis of the so called singularity factor. Numerical examples include the computation of the radiation impedance of vowels /a/, /i/ and /u/ and the input impedance of vowel /a/, for simplified vocal tracts of circular and elliptical cross sections.
\c
\end{abstract}




\maketitle


\section{Introduction} \label{s:I}
\c
\c
\c
Radiation and input impedances of vocal tracts are magnitudes of special interest for voice production. They play a significant role in determining wave radiation from the mouth or in modeling the acoustic coupling between vocal tract and vocal folds acoustics. Impedances can be computed from numerical simulations of vocal tract acoustics. Due to vocal tract intricate geometry, the most extended numerical approach to carry out these simulations is the finite element method (FEM). Several works can be found in literature dealing with FEM computations both in the frequency domain \citep[e.g.,][]{Matsuzaki00,Motoki02,Hannukainen07} and time domain\citep[e.g.,][]{Svancara06,Vampola08a,Vampola11}. Ocasionally other approaches such as finite differences have also been used \citep[e.g.,][]{Takemoto10}.

With regards to voice production, the natural choice turns to be that of working in the time domain. If the wave equation is solved in its mixed form \citep[e.g.,][]{Takemoto10,Codina08}, impedances can be directly computed from the Fourier transforms of the acoustic pressure and acoustic velocity time evolutions. However, this is not possible if the wave equation is solved in irreducible form for the acoustic pressure \citep[e.g.][]{Vampola11} or for the velocity potential \citep[e.g.,][]{Matsuzaki00}. In such cases the acoustic velocity has to be computed from the acoustic pressure or the velocity potential gradients, the convergence error being of higher order than for the primary variables \citep[e.g.,][]{HughesBook}.

In this paper an alternative is proposed to compute vocal tract impedances from time domain simulations without having to compute the acoustic velocity field. The idea is to adapt the experimental two-microphone transfer function method (TMTF) to the numerical framework. Originally developed by \cite{Chung80a}, the simplest version of the experimental TMTF method only requires measuring the time evolution of the acoustic pressure at two points in an impedance duct, and computing the corresponding transfer function. From this transfer function, the radiation and/or input impedances at a given surface can be derived.

Although at first sight applying the TMTF method to compute vocal tract impedances may look straightforward, some issues are worth exploring. These constitute the basis of this work.
\c
On the one hand, we will show the importance of dealing with lossy impedance ducts to reduce the overall time duration of the simulations. The inclusion of wall losses allows to strongly attenuate the duct first eigenmode, which otherwise determines the total duration of the computation. However, this implies using appropriate complex wavenumbers in the TMTF expressions, which are well-known for three-dimensional circular cylindrical ducts, and which will be derived in this work for elliptical cross sectional impedance ducts, given their importance in voice production \citep[see e.g.,][where elliptical vocal tracts are used]{Motoki02,Matsuzaki00}.
\c
On the other hand, the frequency range of validity of the TMTF method will be analyzed. First, it will be shown how the plane wave propagation restriction that limits the TMTF maximum frequency, can be generously surpassed by appropriate location of the virtual microphones (mesh nodes where the acoustic pressure time evolution is collected).
\c
Second, the appropriate virtual microphone spacing will be determined by means of the so called singularity factor (SF) introduced by \cite{Jang98}, according to the maximum frequency of analysis. Throughout the work, time domain FEM simulations for the irreducible wave equation will be performed with an in-house software, to compute vocal tract impedances using the adapted TMTF method. However, any other time domain numerical approach could benefit from the hereafter exposed results. A preliminary version of some of them was presented by the authors in \cite{Arnela12a}.
\c

The paper is organized as follows. Section~\ref{s:II} presents the methodology followed to compute the acoustic impedance of vocal tracts from numerical simulations. In section~\ref{s:III}, the various considerations to be taken into account when adapting the TMTF method to the numerical framework become analyzed. Numerical examples of computed impedances for vocal tracts of circular and elliptical cross section are provided in section~\ref{s:IV}. Finally, conclusions close the paper in section~\ref{s:V}.

\section{Methodology} \label{s:II}

\subsection{The two-microphone transfer function method} \label{ss:TMTF}
\c
A brief survey of the two-microphone transfer function method (TMTF) will be next presented. This method was originally developed by \citet{Chung80a} and later on standardized in the \citet{ISO98} for measuring the normal reflection coefficient of material samples. The specific acoustic impedance $Z$ can be obtained from the latter.
The TMTF proceeds as follows. First, plane waves are generated at the entrance of a duct of length $L$, referred to as the impedance duct. The acoustic pressure signals $P_1 (f)$ and $P_2(f)$ become measured at two points $x_1$ and $x_2$ close to the impedance duct exit, which is designated as the reference surface, and the transfer function
$H_{12}(f) =P_2 (f)/P_1 (f)$ is computed. The reflection coefficient $R_1$ at position $x_1$ is given by 
\c
\begin{align}
  R_1=\frac{H_{12}-H_I}{H_R - H_{12}},
\label{eq:R1}
\end{align}
\c
with $H_I$ and $H_R$ respectively standing for the incident and reflected wave transfer functions. Assuming plane wave propagation, the transfer functions $H_I$ and $H_R$ become $H_I=e^{-jk_z s}$, $H_R=e^{jk_z s}$, with $s=\left|x_1-x_2\right|$ being the distance between microphones and $k_z$ the wavenumber in the axial direction. In order to translate the reflection coefficient to the reference surface, defined at $x=0$, a factor $e^{j 2k_z x_1}$ is introduced in \eqref{eq:R1}, leading to the following expression for the normal reflection coefficient
\c
\begin{align}
  R=R_1 e^{j 2k_z x_1}=\frac{H_{12}-e^{-jk_z s}}{e^{jk_z s} - H_{12}}\ e^{j 2k_z x_1}.
\label{eq:R}
\end{align}
\c
The normalized specific acoustic impedance $Z'$ can be finally obtained by means of
\c
\begin{align}
Z'=Z/Z_0=\frac{1+R}{1-R},
\label{eq:Z}
\end{align}
\c
with $Z$ standing for the specific acoustic impedance and $Z_0=\rho_0 c_0$ for the characteristic impedance.
$Z$ is usually split in its real $R$ (resistive) and imaginary $X$ (reactive) components, $Z = Z_0 (R + j X)$.

\subsection{Problem statement} \label{ss:statement}
\c
\begin{figure} [!t]
	\centering
	\includegraphics[width=75mm]{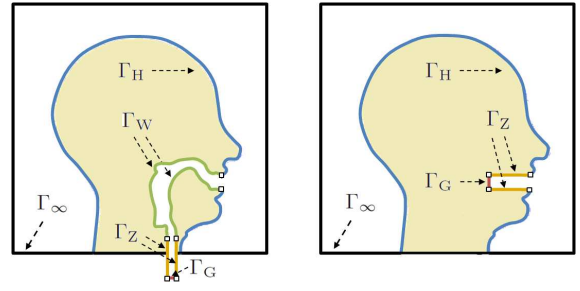}
	\caption{(Color online) A sketch for problem \eqref{eq:WaveEq} in text. Computational domain for the input impedance (left) and for the radiation impedance (right).}
	\label{Fig:ProblemBC}
\end{figure}
\c
Suppose that we want to compute the radiation impedance $Z_r$ or the input impedance $Z_{in}$ of a vocal tract from a time domain numerical simulation using the two-microphone transfer function (TMTF) method. We first need to couple an impedance duct to the reference surface, the duct having the same section and shape than this surface.
\c
For instance, to compute the input impedance of a vocal tract, the impedance duct will be coupled to the vocal tract at the glottal section (see Fig.~\ref{Fig:ProblemBC}), whilst for the radiation impedance, the duct will be coupled to the mouth exit replacing the existing vocal tract geometry (see Fig.~\ref{Fig:ProblemBC}).

The next step consists in carrying out a time domain numerical simulation for the acoustic pressure evolution.
Let us denote by $\Omega$ the finite computational domain with boundary $\partial \Omega$. We can identify on $\partial \Omega$ four non-intersecting regions (see Fig.~\ref{Fig:ProblemBC}):
$\GammaG$ and $\GammaZ$ for the impedance duct boundaries, $\GammaW$ for the vocal tract walls, $\GammaH$ for the human head boundary, and $\GammaI$ for the external boundary, where a fictitious non-reflecting condition has to be imposed. We aim at finding the acoustic pressure field $p(\x,t)$ in $\Omega$ that satisfies
\c
\begin{subequations}
\begin{align}
&\(\partial^2_{tt} - c^2_0\nabla^2 \) p (\x,t)  = 0        &{\rm in}& ~ \Omega,    ~t>0, \label{eq:WaveEqa}
\end{align}
with boundary and initial conditions
\begin{align}
& \nabla p (\x,t) \cdot \n = - \rho_0/S \partial_t Q(t)     &{\rm on}& ~ \GammaG,   ~t>0, \label{eq:WaveEqBC1} \\
& \nabla p (\x,t)\cdot \n = - \mu_w /c_0 \partial_t p (\x,t)  &{\rm on}& ~ \GammaW,  ~t>0, \label{eq:WaveEqBC2} \\
& \nabla p (\x,t)\cdot \n = - \mu_z /c_0 \partial_t p (\x,t)  &{\rm on}& ~ \GammaZ,  ~t>0, \label{eq:WaveEqBC5} \\
& \nabla p (\x,t)\cdot \n = 0                               &{\rm on}& ~ \GammaH, ~t>0, \label{eq:WaveEqBC3} \\
& \nabla p (\x,t)\cdot \n = 1/c_0 \partial_t p (\x,t)       &{\rm on}& ~ \GammaI,   ~t>0, \label{eq:WaveEqBC4} \\
& p(\x,t)=0,\ \partial_t p(\x,t) =0                         &{\rm in}&~\Omega, ~t=0.
\end{align}
\label{eq:WaveEq}
\end{subequations}
\c
In Eq.~\eqref{eq:WaveEq}, $c_0$ denotes the speed of sound, $\rho_0$ the air density, $S$ the impedance duct cross sectional area at $\GammaG$, $\partial_t\equiv\partial / \partial t $ designates the partial time derivative and $\n$ the normal vector pointing outwards a surface. With regards to boundary conditions (BC), $Q(t)$ in \eqref{eq:WaveEqBC1} stands for a volume velocity generated by an imaginary loudspeaker. The BCs in \eqref{eq:WaveEqBC2} and \eqref{eq:WaveEqBC5} account for constant frequency losses at the inner walls, being $\mu$ the boundary admittance coefficient (subindexes $w$ and $z$ simply indicate that different absorption values can be introduced at each boundary). $\mu$ is related to the wall impedance $Z_w$ by means of $\mu= \rho_0 c_0 / Z_w$.
The BC \eqref{eq:WaveEqBC3} expresses that the human head is taken as a rigid surface ($\mu=0$). Finally, BC \eqref{eq:WaveEqBC4} is the well-known Sommerfeld radiation condition, which guarantees that emanating waves from the mouth propagate outwards to infinity. However, this condition is only optimal for sound waves impinging on $\GammaI$ in the normal direction. To overcome this problem, we have made use of a Perfectly Matched Layer (PML) \citep{Berenger94} for the wave equation in its irreducible form. In particular, we have adapted the PML originally developed for the finite difference framework \citep{Grote10} and formulate it for our finite element code. Details on the implemented numerical scheme can be found in appendix~\ref{AppendixA}.

As the simulation evolves, the acoustic pressure signals $p_1(t)$ and $p_2(t)$ are collected at two nodes of the finite element mesh, as if they were virtual microphones.
From their Fourier Transform we obtain $P_1(f)$ and $P_2(f)$, and making use of the TMTF method described in section~\ref{ss:TMTF}, we can compute the acoustic impedances.
\c
\begin{figure}[!t]
  \centering
  \includegraphics[width=76mm]{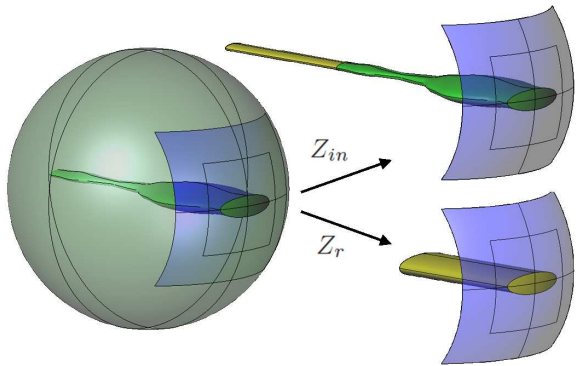}
  \caption{(Color online) Geometries used for computing the input impedance (right top) and the radiation impedance (right bottom) of the elliptical vowel /a/ (left).}
  \label{Fig:Geometry}
\end{figure}
\c

\subsection{Vocal tract models}  \label{ss:VT}
\c
\c
\begin{figure}[!t]
  \centering
  \includegraphics[width=63mm]{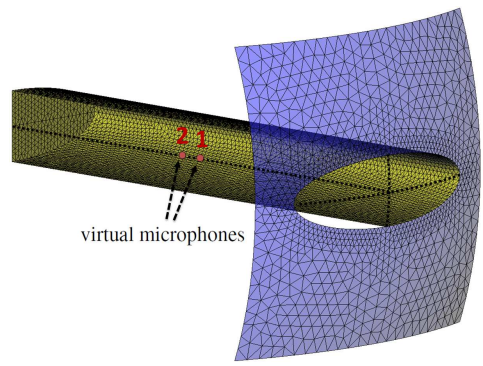}
  \caption{(Color online) Cut with surface mesh details of the impedance duct for the elliptical /a/ radiation impedance computation. Dots indicate locations of virtual microphones for capturing $P_1(f)$ and $P_2(f)$ in text.}
  \label{Fig:Zduct}
\end{figure}
\c
The vocal tracts of the three extreme cardinal vowels /a/, /i/ and /u/ with circular and elliptical cross section have been considered. In order to shorten notation, we will refer to them for instance as circular /a/ or elliptical /a/.
\c
For the circular vocal tracts, we have used the simplified vocal tract geometries generated from sections provided by \cite{Story08}. With regards to the elliptical vocal tracts (see~Fig.~\ref{Fig:Geometry}), we have reshaped the circular sections according to the eccentricity of the elliptical mouth apertures described in \cite{Fromkin64}. A spherical surface of radius $0.09$~m has been used to emulate the human head in all cases.

Following the procedure outlined at the beginning of section~\ref{ss:statement} (see~Fig.~\ref{Fig:Geometry}), we have replaced the vocal tract geometry by an impedance duct with equal mouth aperture to compute the vocal tract radiation impedance, whereas we have coupled an impedance duct at the glottal section of the vocal tract to compute its input impedance. The impedance duct has a length of $L=0.1$~m to fulfill the requirements of the standard \citet{ISO98} (the length should be at least three times the duct radius or the major semi-axis).
\c
The virtual microphones have been located at the centerline of the impedance duct and separated a distance $s=0.01$~m apart (see~Fig.~\ref{Fig:Zduct}). Following the recommendations of the \citet{ISO98}, the first virtual microphone has been placed at a distance from the reference surface slightly larger than two times the impedance duct radius, or the major semi-axis.
\c
Concerning the reference surface for vocal tract radiation impedances computations, it should be noted that it is well-defined for the circular case given that the intersection of a cylindrical vocal tract with a spherical human head results in a flat disk. However, this is not the case if an elliptical vocal tract is used. For such cases we have chosen  the elliptical section where the major semi-axis intersects the sphere as the reference surface (see~Fig.~\ref{Fig:Zduct}).
\c

\subsection{Simulation details}  \label{ss:sim}
\c
The computational domain consists of an outer volume of dimensions $0.25\times 0.2 \times 0.2$~m, where the spherical head has been placed so that sound waves can emanate from the mouth. This volume has been surrounded with a PML of width $0.1$~m to absorb any incident wave. The PML has been configured to get a relative reflection coefficient of $r_\infty=10^{-4}$ (see appendix~\ref{AppendixA}).
\c
The computational domain has been meshed using unstructured tetrahedral elements with a size comprising $h\approx0.1$~cm inside the impedance duct, $h\approx0.5$~cm in the outer volume and $h\approx0.75$~cm in the PML region (see Fig.~\ref{Fig:Zduct} to appreciate some mesh details).

Problem \eqref{eq:WaveEq} with a PML included has been solved using the finite element approach described in appendix~\ref{AppendixA}. A time interval of total duration $T=30$~ms has been simulated using a sampling rate of $f_s=1/\Delta t=2000$~KHz. The values $c_0=345$~m/s and $\rho_0=1.1933~{\rm kg/m^3}$ have respectively been chosen for the speed of sound and for the air density.
\c
Concerning boundary conditions, a wideband impulse has been used for the volume velocity $Q(t)$ in \eqref{eq:WaveEqBC1}, and input at the impedance duct entrance. We have used a gaussian pulse of the type \citep{Takemoto10}
\c
\begin{equation}
  gp(n)=e^{\[ (\Delta t ~ n - T_{gp}) 0.29 T_{gp}   \]^2} {\rm [m^3/s]},
\end{equation}
\c
with $T_{gp}=0.646/f_0$ and $f_0=10$~KHz. To avoid numerical errors beyond the maximum frequency of interest ($f_{max}=10$~KHz), this pulse has been filtered using a low-pass filter with cutoff frequency $10$~KHz.
\c
For the boundary admittance coefficient at the vocal tract walls we have used $\mu_w=0.005$, which corresponds to the wall impedance of the vocal tract tissue $Z_{w}=83666$~${\rm Kg/m^2s}$ \citep[see][]{Svancara06}. For the impedance duct, we have used the artificial value $\mu_z=0.01$.

\section{The two-microphone transfer function method for numerical simulations} \label{s:III}

\subsection{Damping the first impedance duct eigenmode} \label{ss:Zw}
\c
Although theoretically numerical simulations to compute vocal tract impedances could be carried out using a lossless impedance duct, including boundary losses is mandatory for the simulations to have a reasonable duration. From an experimental point of view, time duration is not a problem given that, for example, a measurement that lasts 5 seconds can be easily performed. However, in the numerical framework the CFL stability condition pose severe restrictions on the time step $\Delta t$ to be used, so that for intricate and large computational domains, a 5 seconds event may involve several hours of computational time.

\begin{figure}[!tp]
  \centering
  \includegraphics[height=54mm]{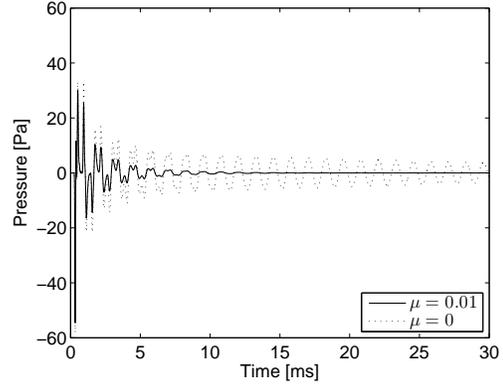}
  \caption{Acoustic pressure evolution for circular /a/ at virtual microphone $\# 1$ for the lossless ($\mu$=0) and lossy ($\mu$=0.01) cases.}
  \label{Fig:PressureAb}
\end{figure}

\begin{figure*}[!t]
  \centering
  \subfloat[$\mu=0$ \label{Fig:SpectrogramAb0}]  {\includegraphics[height=54mm]{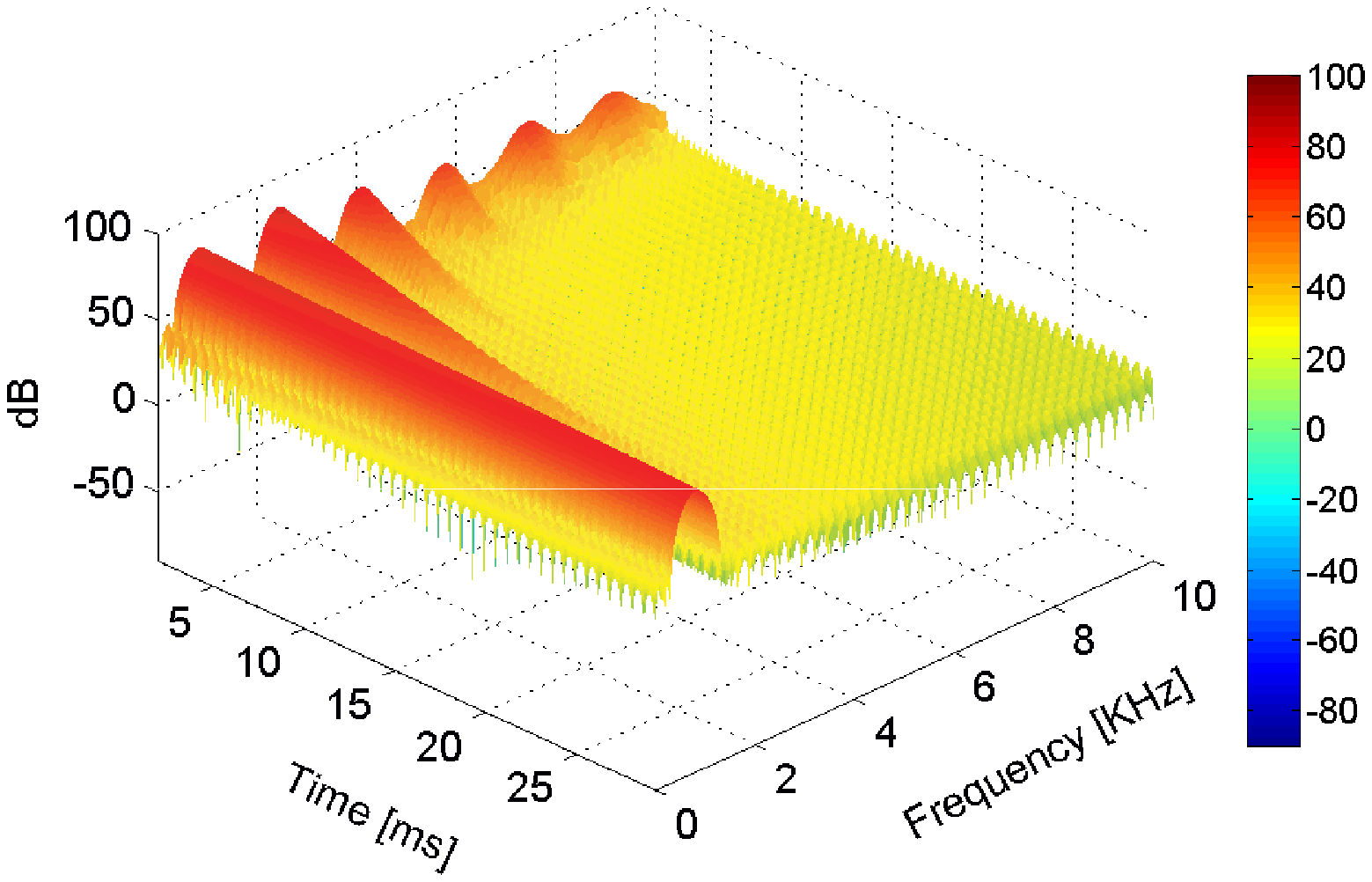}}\hspace{1cm}
  \subfloat[$\mu=0.01$ \label{Fig:SpectrogramAb}] {\includegraphics[height=54mm]{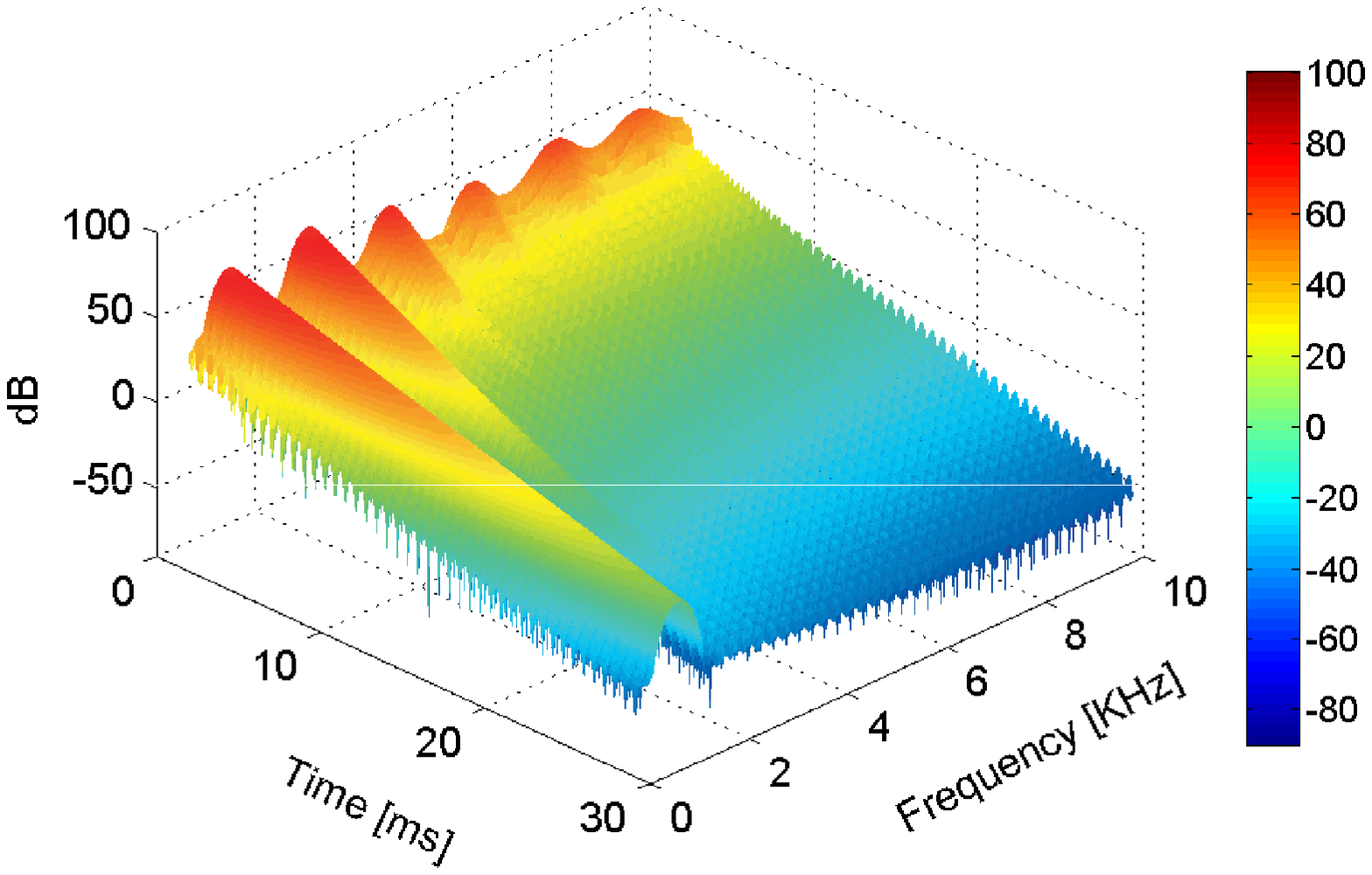}}
  \caption{(Color online) Spectrograms for the acoustic pressures in Fig.~\ref{Fig:PressureAb}: (a) lossless impedance duct ($\mu$=0) and (b) lossy impedance duct ($\mu$=0.01).}
  \label{Fig:Spectrogram}
\end{figure*}

Consider for example, the radiation impedance computation of the circular /a/ in two cases: i) a lossless impedance duct with $\mu=0$ and ii) a lossy impedance duct with $\mu=0.01$. Let us first focus on the acoustic pressure collected at the first virtual microphone $\# 1$ (see Fig.~\ref{Fig:Zduct}) and plot its time evolution for the lossless and lossy cases in Fig.~\ref{Fig:PressureAb}.
\c
\c
For the former, the $30$~ms duration of the simulation has not sufficed to attenuate the signal acoustic pressure inside the impedance duct, whilst it has decayed in about 15~ms for the lossy case. It should be remarked that given that the acoustic pressure has to be Fourier transformed to apply the TMTF method, it is necessary for it to vanish to zero during the simulation interval, to avoid spurious errors in this operation.
\c
Some more insight on what is going on inside the impedance duct can be obtained from the spectrograms (time vs frequency) of the acoustic pressure signals for the lossless and lossy cases (see Fig.~\ref{Fig:Spectrogram}).
\c
\c
As observed, the problem arises from the difficulty to attenuate the first duct eigenmode. This is attributed to the fact that radiation is by far a more effective energy dissipating mechanism at high frequencies than at low frequencies. Therefore, the inclusion of wall damping clearly helps to overcome this problem and noticeably shortens the duration of the simulation.

\subsection{Analytical expressions for the complex axial wavenumber} \label{ss:kz}
\c
Introducing artificial damping at the impedance duct walls of the numerical model involves dealing with complex axial wavenumbers~$k_z$ in expression \eqref{eq:R} of the TMTF method. In experimental measurements, a calibration procedure is usually conducted for estimating the attenuation factor \citep[e.g.][]{Boonen09}. In the numerical case, analytical expressions can be deduced for~$k_z$. We will next show how to relate $k_z$ with the admittance boundary coefficient $\mu_z$ in problem \eqref{eq:WaveEq}, for three-dimensional ducts of elliptical cross section.

To proceed, it is convenient to express the wave equation (read also the Helmholtz equation) in elliptic cylindrical coordinates $(\xi,\eta,z)$ \citep[see e.g.,][]{Lowson75}. For each constant value of $z$, the coordinate lines correspond to confocal ellipses and hyperbolae. Curves of constant $\xi$ are ellipses whilst curves of constant $\eta$ are hyperbolae. Using this system of coordinates, the boundary condition \eqref{eq:WaveEqBC5} at the duct wall ($\xi=\xi_0$) can be expressed as \citep[see e.g.,][]{Oliveira10}
\c
\begin{equation}
  \frac{\partial \hat{p}}{\partial \xi}+j\mu_z \kappa \sqrt{1-e^2\cos^2\eta}\ \hat{p} =0\ \ \ {\rm at}\ \xi=\xi_0,
\label{eq:BCee}
\end{equation}
\c
where $e=\sqrt{1-b_e^2/a_e^2}$ is the eccentricity of the ellipse defining the duct boundary at $\xi_0$, which has focal distance $f=a_e e$ and major and minor semi-axes $a_e=f\cosh \xi_0$ and $b_e=f\sinh \xi_0$. $\kappa \equiv k_0 a_e$ is the reduced (adimensional) wavenumber.
To simplify notation, use will be also made of the parameter
\c
\begin{equation}
q \equiv \left(\frac{k_\bot f}{2}\right)^2=\left(\frac{k_\bot a_e e}{2}\right)^2,
\label{eq:q}
\end{equation}
\c
with $k_\bot \equiv \sqrt{k_0^2-k_z^2}$ standing for the transverse wavenumber.

\c
Separation of variables for the Helmholtz equation in elliptic cylindrical coordinates results in the so-called Mathieu radial and angular equations. The solutions need to be $2\pi$ periodic in $\eta$, which plays the role somehow analogous to $\theta$ for the circular case \citep[developed e.g. in][]{Munjal87}. The periodic solutions are given by products of cosine elliptic functions $ce_m(\eta,q)$, which are even, with the radial Mathieu functions $Je_m(\xi,q)$ related to them, and by sine elliptic functions $se_m(\eta,q)$, which are odd, and their corresponding radial Mathieu functions $Jo_m(\xi,q)$ \citep[see][]{Gutierrez-Vega00}. However, no linear combination of $ce_m(\eta,q)Je_m(\xi,q)$ and $se_m(\eta,q)Jo_m(\xi,q)$ is allowed since the sets of characteristic values for $ce_m(\eta,q)$ and $se_m(\eta,q)$ are different \citep[the situation is somehow similar to what occurs for rectangular ducts, see e.g.,][]{Munjal87}. Given that we are interested in the case of plane wave propagation inside the duct, we will be interested in the lowest modal indices and have to deal with the even case.

The even radial Mathieu functions $Je_m(\xi,q)$ admit a series factorization in terms of Bessel functions. The factorization depends on $m$ being even or odd so we can get even-even $Je_{2k}(\xi,q)$ and even-odd $Je_{2k+1}(\xi,q)$ radial Mathieu function developments. As we want to consider the case $m=0$ we have to deal with the former. It turns that \citep[][]{Gutierrez-Vega00}
\begin{align}
Je_{2k}(\xi,q)&=\frac{ce_{2k}(0,q)}{A_0}\sum_{j=0}^{\infty} A_{2j}J_{2j}(2\sqrt{q}\sinh\xi)
\label{eq:serpmat}
\end{align}
with $J_m$ corresponding to the Bessel function of order $m$. Taking $k=j=0$ we get
\c
\begin{align}
\hat{p}(\xi,\eta,z)&\simeq ce_0(\eta,q) ce_{0}(0,q) J_0 \left(2\sqrt{q}\sinh\xi\right), \label{eq:pmat}
\end{align}
\c
with derivative
\c
\begin{align}
\frac{\partial \hat{p}(\xi,\eta,z)}{\partial \xi} &\simeq - ce_0(\eta,q) ce_{0}(0,q) \nonumber \\
&\times 2\sqrt{q}\cosh\xi J_1 \left(2\sqrt{q}\sinh\xi\right).
\label{eq:dpmat}
\end{align}
\c
Substituting \eqref{eq:pmat} and \eqref{eq:dpmat} into the boundary condition \eqref{eq:BCee} and evaluating at $\xi=\xi_0$ we get
\c
\begin{equation}
 \frac{2\sqrt{q}\cosh\xi_0 J_1 \left(2\sqrt{q}\sinh\xi_0\right)}{\kappa J_0 \left(2\sqrt{q}\sinh\xi_0\right)}=j\mu_z  \sqrt{1-e^2\cos^2\eta}.
\label{eq:cel1}
\end{equation}
\c
In order to eliminate the $\eta$ dependence in the r.h.s of \eqref{eq:cel1} we can integrate at both sides from $0$ to $\pi/2$. Changing variables to $v=\cos \eta$ shows that the integral in the r.h.s is a complete elliptic integral of the second kind, whose solution can be expressed as a series in terms of even powers of the eccentricity \citep[see][]{Abramowitz70}
\c
\begin{align}
&\int_0^{\pi/2}\sqrt{1-e^2\cos^2\eta} d\eta=\int_0^1 \frac{\sqrt{1-e^2v^2}}{\sqrt{1-v^2}}dv \nonumber \\
&= \frac{\pi}{2}\left[1-\sum_{n=1}^\infty\left(\frac{(2n-1)!!}{(2n)!!}\right)^2\frac{e^{2n}}{2n-1}\right]\equiv I(e) \label{eq:EI}
\end{align}
\c
Using this result, approximating the Bessel functions to first order, $J_0(x)\simeq 1$, $J_1(x)\simeq x/2$, and considering the notation introduced above, it follows from Eq.~\eqref{eq:cel1} that
\c
\begin{align}
q=\frac{j\mu_z \kappa I(e)}{\pi \cosh\xi_0 \sinh \xi_0}=j\frac{\mu_z k_0 I(e) a_e^2 e^2}{\pi b_e}. \label{eq:cel2}
\end{align}
\c
From \eqref{eq:q} we then get
\c
\begin{align}
k_{\bot}^2=j\frac{ 4 \mu_z k_0 I(e)}{\pi b_e} \label{eq:cel3}
\end{align}
\c
and the axial wavenumber that we were looking for becomes
\c
\begin{align}
  k_z=\sqrt{k^2_0 - k^2_\bot}
  \simeq k_0 \sqrt{1-j\frac{4 \mu_z I(e)}{k_0 \pi b_e}}.
\label{eq:kz3Elcyl}
\end{align}
\c
\begin{figure*}[!t]
  \centering
  \subfloat[\label{Fig:ZiE10nokz}] {\includegraphics[height=54mm]{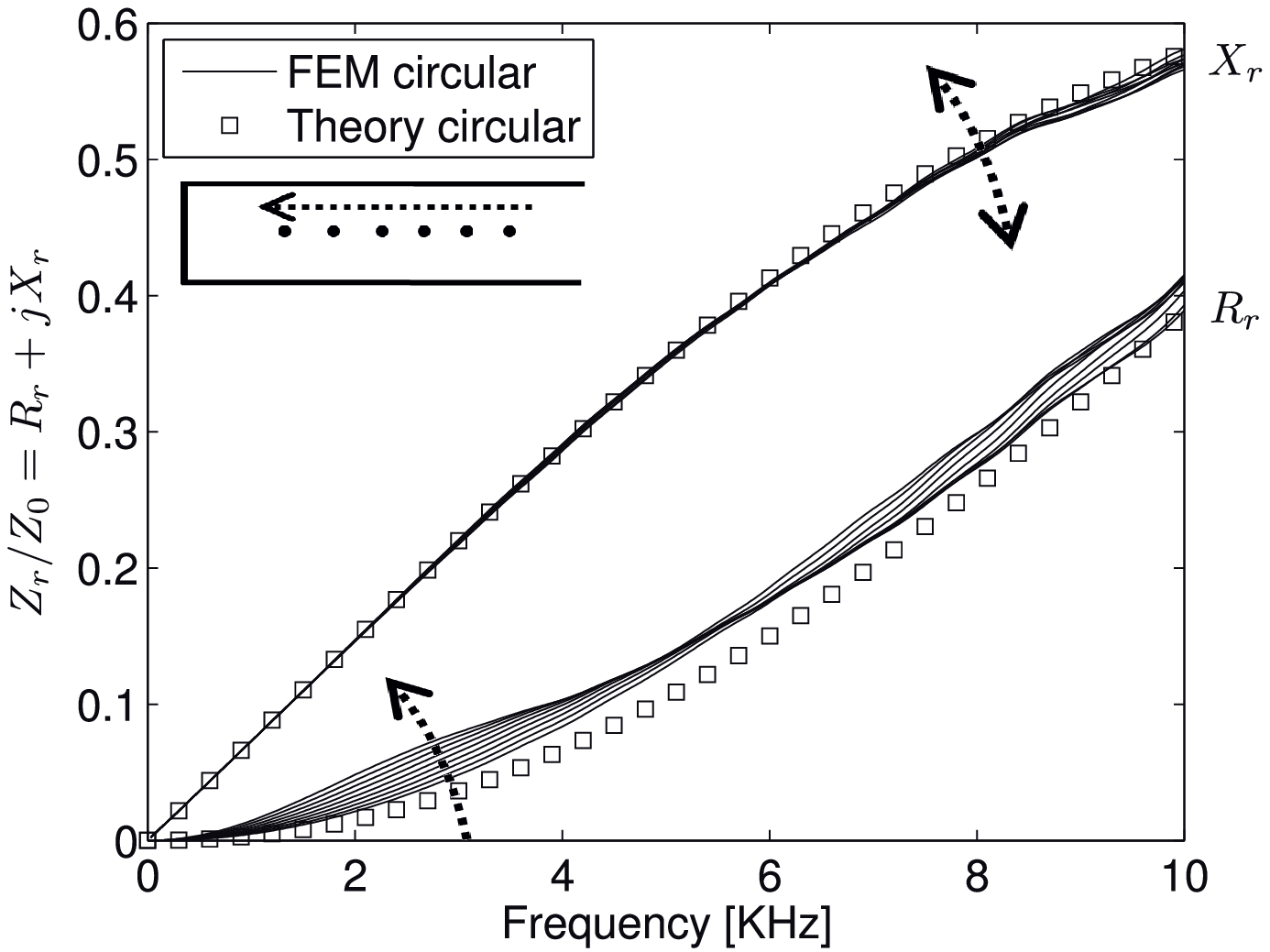}}\hspace{1cm}
  \subfloat[\label{Fig:ZiE10kz}] {\includegraphics[height=54mm]{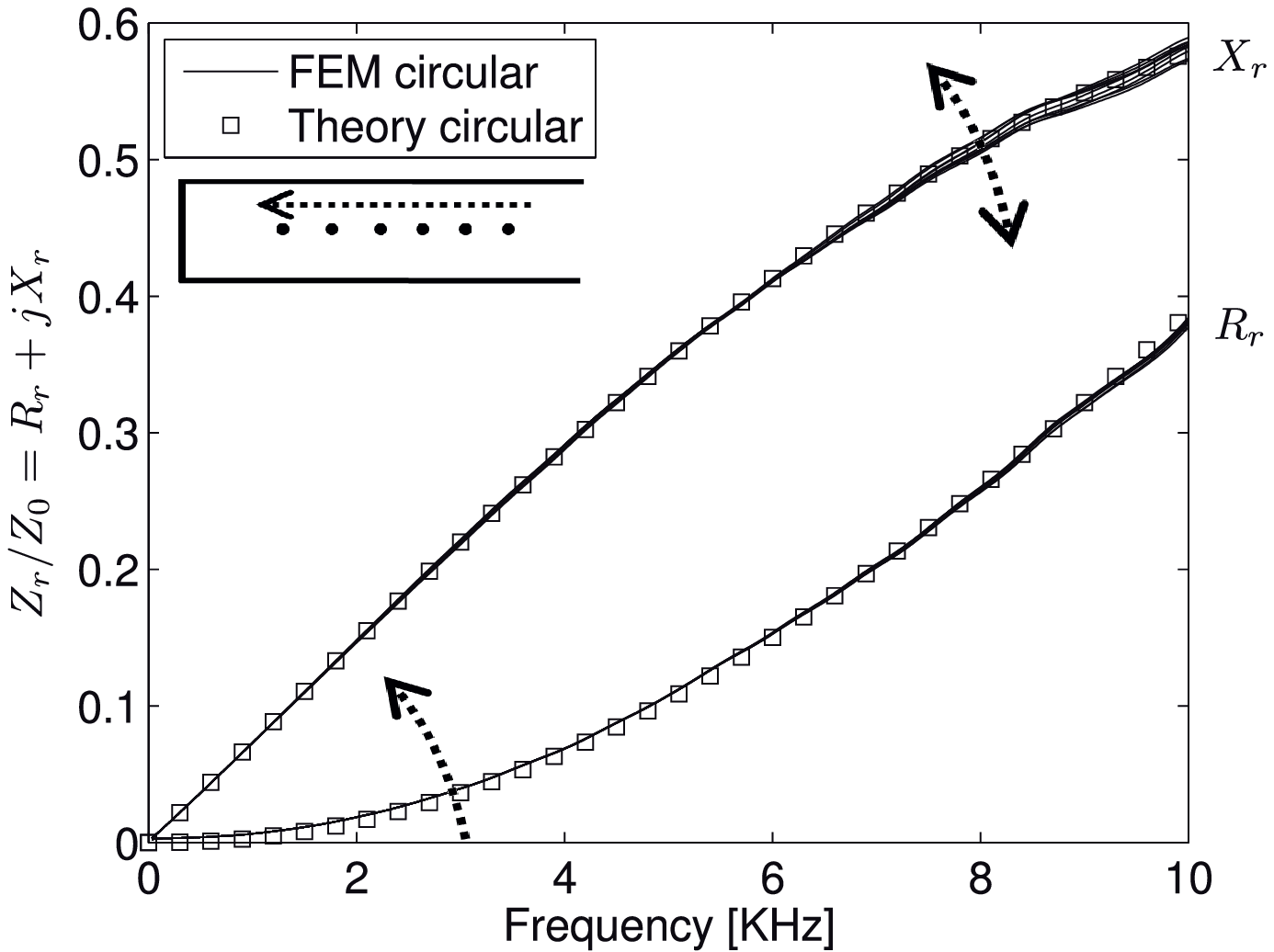}} \\
  \subfloat[\label{Fig:ZiE540nokz}] {\includegraphics[height=54mm]{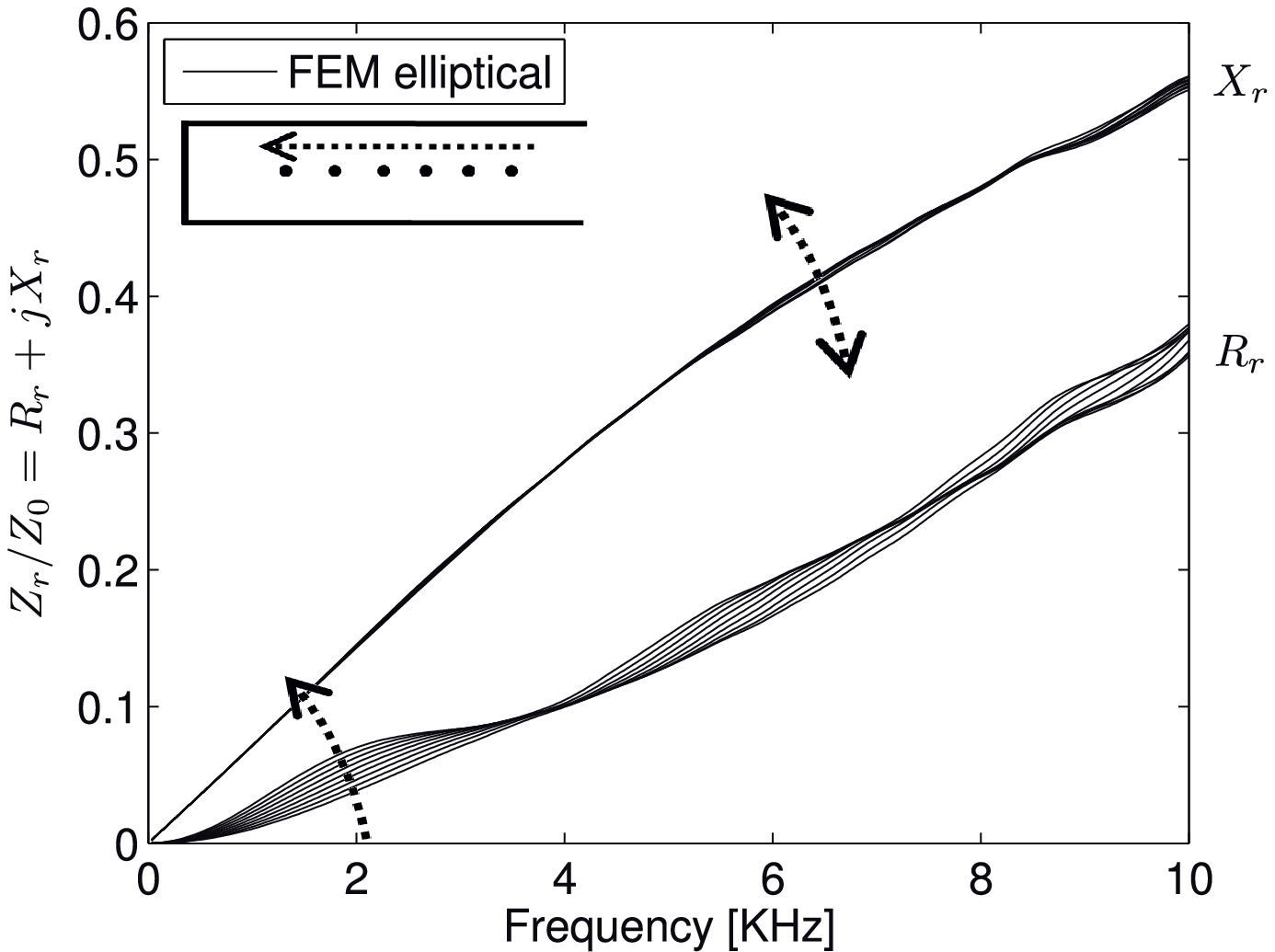}}\hspace{1cm}
  \subfloat[\label{Fig:ZiE540kz}] {\includegraphics[height=54mm]{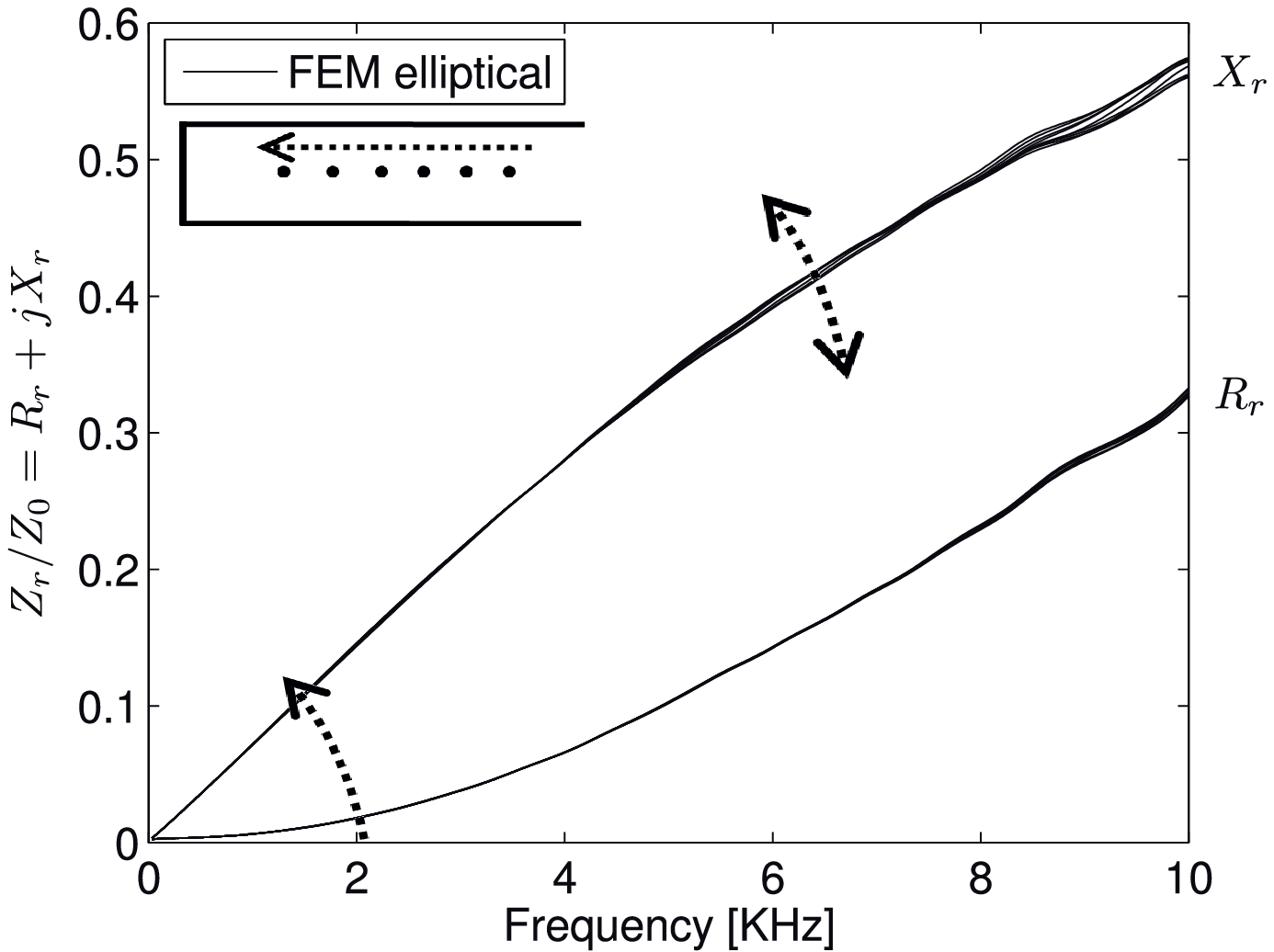}} \\
  \caption{Radiation resistance $R_r$ and reactance $X_r$ for vowel /i/ with circular (top) and elliptical (bottom) mouth aperture for different pairs of virtual microphones. (a)(c) Real wavenumber computations. (b)(d) Complex wavenumber computations. Solid lines correspond to FEM simulations and square symbols to the theoretical model, only available for a circular piston.}
  \label{Fig:Zikz}
\end{figure*}
\c
In the case of zero eccentricity we get $I(0)=\pi/2$, $a_e=b_e\equiv a$, and recover the axial wavenumber for the circular case
\c
\begin{equation}
  k_{z}=\sqrt{k^2_0 - k^2_{r}}\simeq k_0 \sqrt{1-j\frac{2 \mu_z}{k_0 a}},
\label{eq:kz3Dcyl}
\end{equation}
which can be derived from first order approximations in the characteristic equation for circular cylindrical ducts in \cite{Munjal87}.

On the other hand, note that to compute the impedance of vocal tracts, impedance ducts of different cross sections and shapes will be necessary. This implies that the total area of the impedance duct walls will be different in each case, so that if we use the same boundary admittance coefficient $\mu_z$ we will get different internal dissipation.
\c
To guarantee the same amount of absorption, we must ensure that the axial wavenumber remains the same for all cases. Suppose that we want to get the same dissipation in an elliptical impedance duct $i$ than in another elliptical impedance duct $j$, with respective eccentricities $e_i$, $e_j$, and semi-minor axes $b_{e,i}$, $b_{e,j}$. Equating their axial wavenumbers \eqref{eq:kz3Elcyl} we can compute $\mu_{z,i}$ from $\mu_{z,j}$ as
\c
\begin {equation}
  \mu_{z,i} = \mu_{z,j} \frac{I(e_j)}{I(e_i)} \frac{b_{e,i}}{b_{e,j}}.
  \label{eq:muElliElli}
\end {equation}
\c

As mentioned in section~\ref{ss:Zw}, in this work a boundary admittance coefficient of $\mu_z=0.01$ has been chosen for the circular vowel /a/. Therefore, to introduce the same amount of dissipation in all simulations, we have made use of~\eqref{eq:muElliElli}, with $j=$/a/ and $i=$/i/,/u/.
\c
\c
\c
\c
\c

\subsection{Accuracy of complex axial wavenumbers}
\c
Several assumptions have been made in order to derive the axial complex wavenumbers in~\eqref{eq:kz3Elcyl} and~\eqref{eq:kz3Dcyl}. Let us next check if these simple expressions  are precise enough for vocal tract impedance computations and what would happen if real wavenumbers were considered instead. To do so, we have computed the radiation impedance of the circular and elliptical /i/ using complex and real wavenumbers, and considered different pairs of virtual microphones, located at different distances from the duct exit. The computed radiation impedances, splitted in terms of resistance $R_r$ and reactance $X_r$, have been plotted in Fig.~\ref{Fig:Zikz} and compared to the theoretical model of the baffle set in a spherical surface \citep{Morse68}, for the circular case.
\c
\c
\c
The single point arrow on the $R_r$ curves indicates their tendency when the selected pairs of microphones used for the computations are moved away from the duct exit (see duct scheme in the upper left side of figures). The double point arrow indicates that the $X_r$ curves do not follow any clear tendency when changing the pairs of virtual microphones.

If we have a look at Eq.~\eqref{eq:R}, we observe that the term $e^{j 2 k_z x_1}$ is used to translate the reflection coefficient $R_1$ to the reference plane ($x_1$ is the distance from the first virtual microphone to the reference plane).
\c
Taking into account the first order approximation $(1+x)^{1/2}\sim1+ x/2$ in the expression for the complex wavenumber in \eqref{eq:kz3Elcyl} (which is valid for hard wall behavior $\left|Z_w\right|\gg \rho_0 c_0$), it follows that $k_{z,00}\equiv k_z \simeq k_0-j2\mu_z I(e) / (\pi b_e)$. Substituting into the propagator $e^{j 2 k_z x_1}$ results in $e^{j 2 x_1 k_0} e^{j 2 x_1 \mu_z I(e) / (\pi b_e)}$, so that the factor $e^{j 2 x_1 \mu_z I(e) / (\pi b_e)}$ will be missing if a real wavenumber is used. Note that the higher the value of $x_1$ the larger will be the error. This can be appreciated in Fig.~\ref{Fig:ZiE10nokz} for the circular case, where $R_r$ clearly departs from the theoretical curve as $x_1$ increases. As opposite, the error is inexistent when the appropriate complex wavenumber of equation~\eqref{eq:kz3Dcyl} is used in~\eqref{eq:R}, and the computed curve perfectly matches the theoretical one (see Fig.~\ref{Fig:ZiE10kz}). The reactance $X_r$ seems not to be as much  affected as the resistance $R_r$ by the error. 

A very similar behavior can be observed for the elliptical case as shown in Fig.~\ref{Fig:ZiE540nokz} and ~Fig.~\ref{Fig:ZiE540kz}.
\c
Besides, note that both for the circular and elliptical cases, some discrepancies are found for the reactance values at the high frequency range, even when using the correct complex wavenumbers. This is attributed to numerical errors in the simulations to be discussed in section~\ref{ss:SF}.
\c
\c

\subsection{Plane wave propagation restriction} \label{ss:planewave}
\c
One of the main frequency limitations of the TMTF method is that of plane wave propagation assumption.
Next we will examine how this condition can be relaxed for vocal tract impedance ducts with circular and elliptical sections.

\begin{figure}[!h]
  \centering
  \subfloat[\label{Fig:ModesCirc}]{\includegraphics[height=36mm]{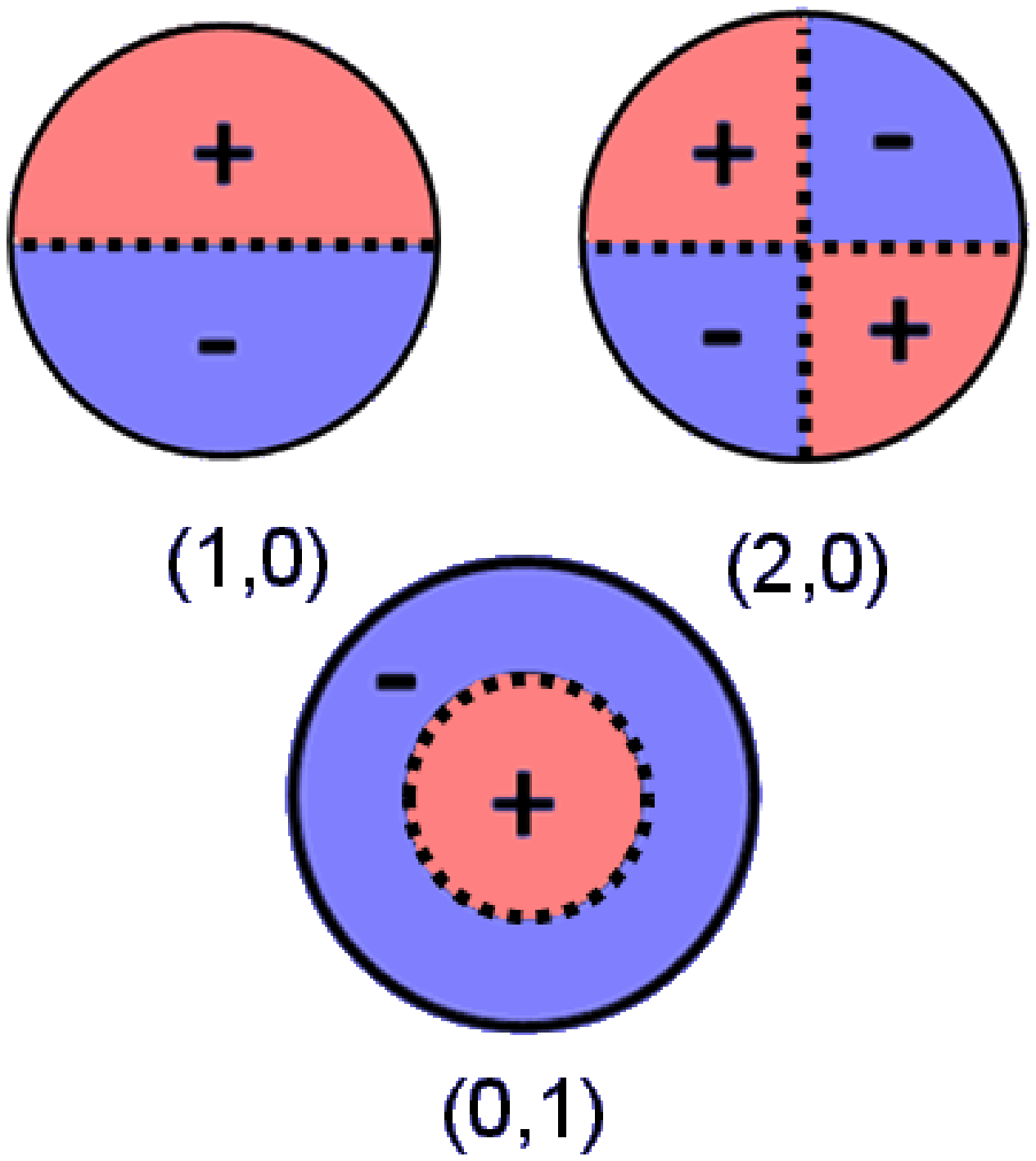}}\hspace{0.5cm}
  \subfloat[\label{Fig:ModesElli}]{\includegraphics[height=36mm]{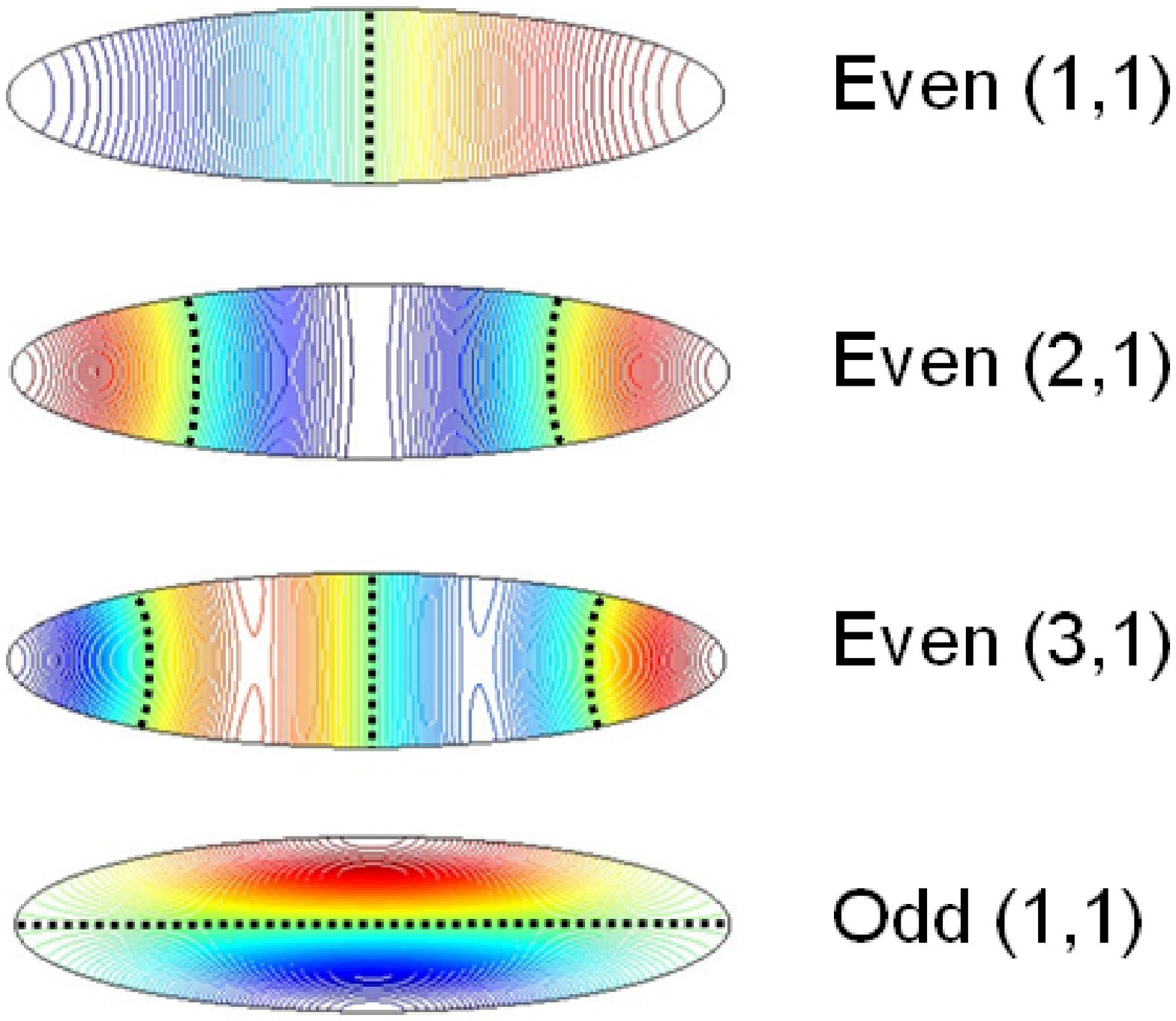}}
  \caption{(Color online) Plots of the lower standing modes in an impedance duct with (a) circular and (b) elliptical section. Discontinuous lines represent nodal planes.}
  \label{Fig:Modes}
\end{figure}

For a circular impedance duct of radius $a$, the first non-planar eigenmode is the (1,0) mode, with a cutoff frequency $f_{(1,0)}=1.84 c_0 / (2 \pi a)$ \citep{Fletcher88}. The next one, is the (2,0) mode, with $f_{(2,0)}=3.05 c_0 / (2 \pi a)$, followed by the (0,1) mode, with $f_{(0,1)}=3.80 c_0 / (2 \pi a)$.
\c
In the experimental framework, the (1,0) mode limits the working frequency range of the TMTF method. However, in the numerical framework, this limitation can be overcome by a proper location of the virtual microphones. If we have a look at the pressure distribution of the first three eigenmodes (see Fig.~\ref{Fig:ModesCirc}), we can observe nodal planes (corresponding to lines in the duct sections of the figure) that cross the centerline of the duct for the (1,0) and (2,0) modes. Consequently, if we precisely locate the virtual microphones at the centerline, the pressure there will not be affected by these modes, but will be only due to plane wave propagation. The TMTF method will be then still applicable at these frequencies and limited by the (0,1) mode, which is the first eigenmode that does not have a nodal line at the center of the duct (see Fig.~\ref{Fig:ModesCirc}).
\c
\c
\c

With regards to elliptical ducts, the frequency range can also be extended. In Fig.~\ref{Fig:ModesElli} we present the pressure patterns of the first four eigenmodes for the elliptical /a/ impedance duct \cite[these can be computed e.g., from the formulas in][]{Oliveira10}. The first non-planar mode is the even mode (1,1). This mode has a nodal plane in the center, so that locating again the virtual microphones in the centerline of the impedance duct will allow to extend the analysis up to the frequency of the even mode (2,1), which is the limiting one in this case.
\c
\begin{table}[!ht]
\caption{Frequency values in KHz for the first eigenmodes of the circular and elliptical impedance ducts used to compute the radiation impedance $Z_r$ and the input impedance $Z_{in}$ for vowels /a/, /i/ and /u/. Stars denote maximum frequency of analysis when the virtual microphones are located at the centerline of the impedance duct.}
\label{tab:modes}
 \setlength{\tabcolsep}{2.2pt}
 \renewcommand{\arraystretch}{1.2}
 \begin {tabular} {c c c c c c c c c c}
  \hline
  \hline
                                      & &  \multicolumn{3}{c}{Circular} &&  \multicolumn{4}{c}{Elliptical} \\
  \cline{3-5} \cline{7-10}
                                      & & (0,1) & (1,0) & (1,1) && E(1,1) & E(2,1) & E(3,1) & O(1,1) \\
  \hline
           & \multicolumn{1}{c}{$/a/$}  & 8.21  & 13.62 & 17*    && 4.68  & 8.6*   & 12.47 & 13.79 \\
  $Z_r$    & \multicolumn{1}{c}{$/i/$}  & 18.8  & 31.13 & 38.8*  && 8.27  & 15.27* & 22.18 & 39.3  \\
           & \multicolumn{1}{c}{$/u/$}  & 44.77 & 74.2  & 92.46* && 22.23 & 40.98* & 59.49 & 84.2  \\
  \hline
  $Z_{in}$ & \multicolumn{1}{c}{$/a/$}  & 23.72 & 39.32 & 48.99* && 13.57 & 24.97* & 36.2  & 40.03  \\
  \hline
  \hline
 \end{tabular}
\end{table}
\c

In Table~\ref{tab:modes} we have computed the first modes of the impedance ducts used in this work, with stars denoting in each case the first mode without a nodal plane at its center (i.e., the limiting one). As expected, the impedance duct with a strongest restriction is that of vowel /a/ (largest mouth aperture), whereas vowel /u/ (smallest mouth aperture) presents the less stringent condition. It can also be observed that working with elliptical mouth apertures results in more restrictive frequency ranges. Note however that by locating the virtual microphones at the centerline of the impedance duct, the upper frequency limit of the experimental TMTF method has been almost increased by a factor~$\sim 2$. Except for the elliptical /a/, the desired frequency range of analysis that goes up to $f_{max}=10$~KHz can thus be attained.
\c
\c
\c

\subsection{The singularity of the TMTF method} \label{ss:SF}
\c
It is well-known that a singularity occurs in the experimental TMTF method when half the wavelength of the acoustic pressure equals the microphone spacing $s$, or one of its multiples. The minimum frequency that satisfies this condition is termed the \emph{critical frequency}, $f_{cr}=c_0 /(2s)$, and imposes a high frequency limit $f_u<f_{cr}$ to the method. On the other hand, a minimum low frequency limit $f_d$ also exists, because the pressure differences measured by the two microphones will be negligible if their distance apart is very small compared to the measured wavelength.
Working close to both limits is not recommended and a suitable option is to take $f_{d} = 0.2f_{cr}<f<0.8f_{cr}\ = f_{u}$ \cite[as proposed in ][]{Jang98}.
\c
\begin{figure}[t]
  \centering
  \includegraphics[height=54mm]{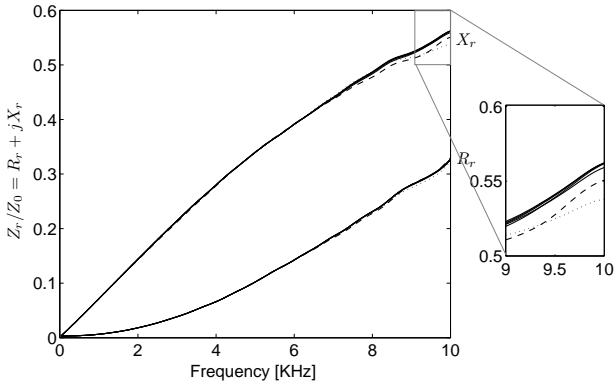}
  \caption{Resistance and reactance of the elliptical /i/ for different microphone spacings $s$. The dashed line stands for $s=0.1$~cm (equal to the mesh size $h$), the dotted line for $s=1.5$~cm (close to the singularity), and solid lines correspond to intermediate values of $s$.}
  \label{Fig:Zr3Ds}
\end{figure}
\c

The above situation contrasts with that of numerical simulations. Suppose that we are interested in making a simulation up to a given frequency $f_{max}$ with corresponding wavelength $\lambda_{min}$. If we had unlimited computational resources, so that we could work with a fine enough mesh to meet the standard criterion of 10 nodes per wavelength for $\lambda_{min}$ \citep[see e.g.,][]{Ihlenburg98}, numerical errors would be negligible. In such situation the only frequency upper limitation would come from the plane wave restriction discussed in section~\ref{ss:planewave}, and we could take e.g., $f_d = 0<f<f_{max}=f_{cr}=f_{u}$ as the operational working frequency range. This expression together with $f_{cr}=c_0 /(2s)$ provides the following possible values for the virtual microphone spacing $s$
\c
\begin{equation}
h \leq s<0.5\lambda_{min}.
\label{eq:sh}
\end{equation}
\c
Note from \eqref{eq:sh}, that obviously the microphone spacing $s$ cannot  be smaller than the mesh size $h$.

However, working with very fine meshes can be unfeasible in many simulations and the 10 nodes per wavelength criterion is often sacrificed to lower the computational cost. This results in the appearance of some numerical errors at high frequencies as happens with the simulations performed throughout this work, where $34.5$, $6.9$ and $4.6$ nodes per $\lambda_{min}$ have been respectively taken within the impedance duct, the outer volume and the PML region (see section~\ref{ss:sim} for mesh details). In Fig.~\ref{Fig:Zr3Ds} we have plotted the radiation resistance and reactance for the elliptical /i/ computed using different microphone spacings, $s=0.1$, $0.25$, $0.5$, $0.75$, $1$, $1.25$, $1.5$~cm, which respectively yield $s/\lambda_{min}\sim0.03$, $0.07$, $0.15$, $0.22$, $0.29$, $0.36$, $0.43$. The first spacing ($s=0.1$~cm) has been chosen equal to the mesh size $h$ inside the impedance duct (dashed curve in Fig.~\ref{Fig:Zr3Ds}), whilst the last one ($s=1.5$~cm) has a value close to the singularity value $0.5\lambda_{min}$ (dotted curve in Fig.~\ref{Fig:Zr3Ds}).
\c
No differences can be appreciated in Fig.~\ref{Fig:Zr3Ds} between the various $R_r$ and $X_r$ curves for the low-mid frequency range as there are no significant numerical errors at these frequencies for the present simulations. Slight differences only become apparent for frequencies higher than 8 KHz (see zoom in Fig.~\ref{Fig:Zr3Ds}), the largest ones precisely corresponding to the reactances computed with the limiting values of $s=0.1$~cm (mesh size) and $s=1.5$~cm (singularity value).
\c
\c
\begin{figure}[!t]
  \centering
  \includegraphics[height=54mm]{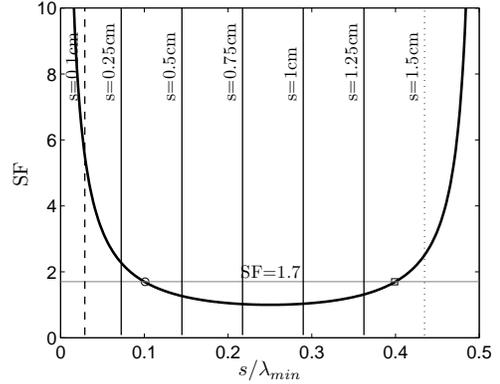}
  \caption{Singularity factor SF of the TMTF method for a lossless impedance duct. The circle and the square denote the low and high limits for the microphone spacing $s$ obtained when ${\rm SF}=1.7$. Vertical lines correspond to the microphone spacing configurations used in the example of Fig.~\ref{Fig:Zr3Ds}.}
  \label{Fig:Zr3DSF}
\end{figure}
\c
\c
\c
\begin{figure*}[!t]
  \centering
  \subfloat[\label{Fig:Rvowels}] {\includegraphics[height=54mm]{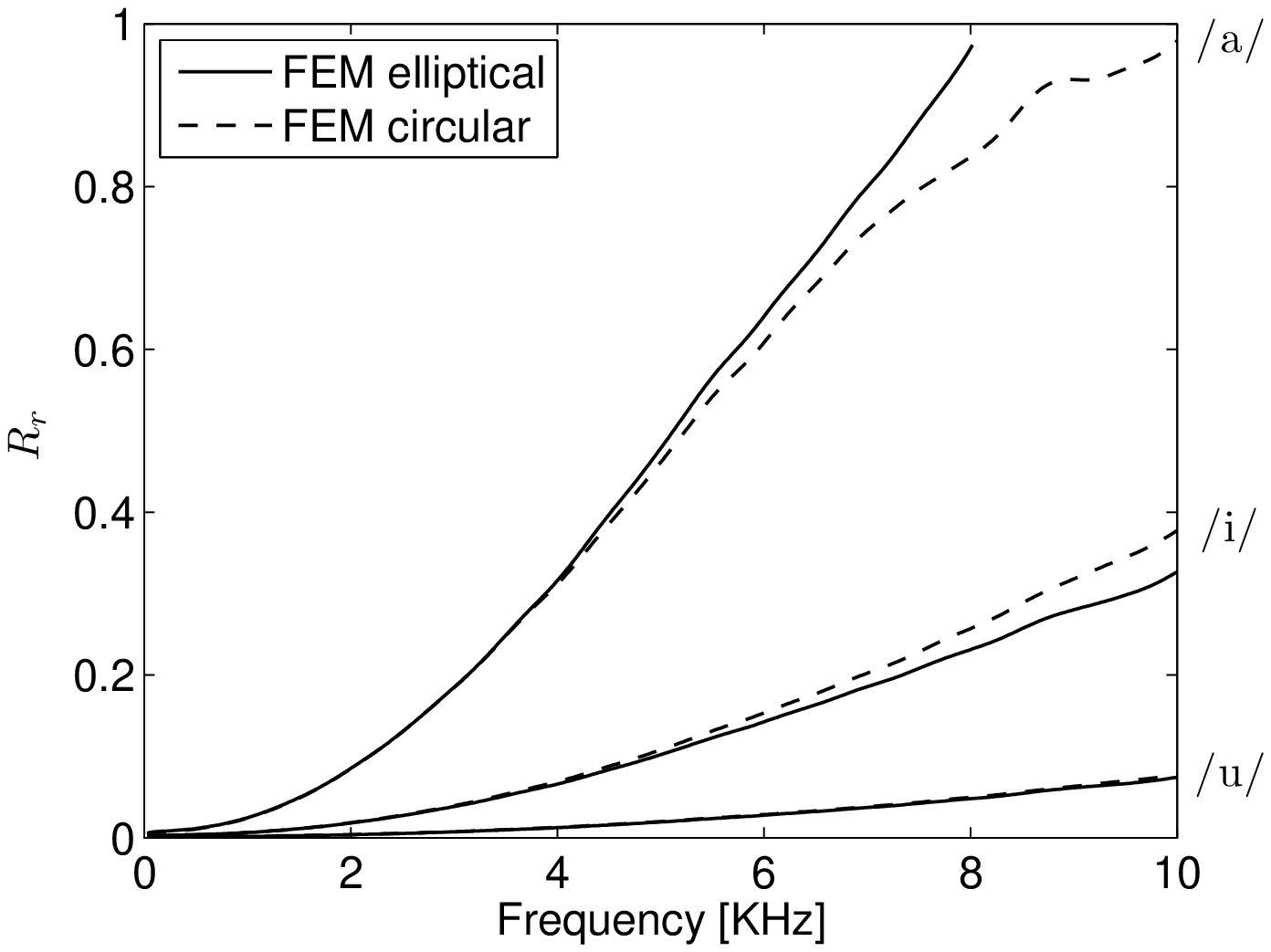}}\hspace{1cm}
  \subfloat[\label{Fig:Xvowels}] {\includegraphics[height=54mm]{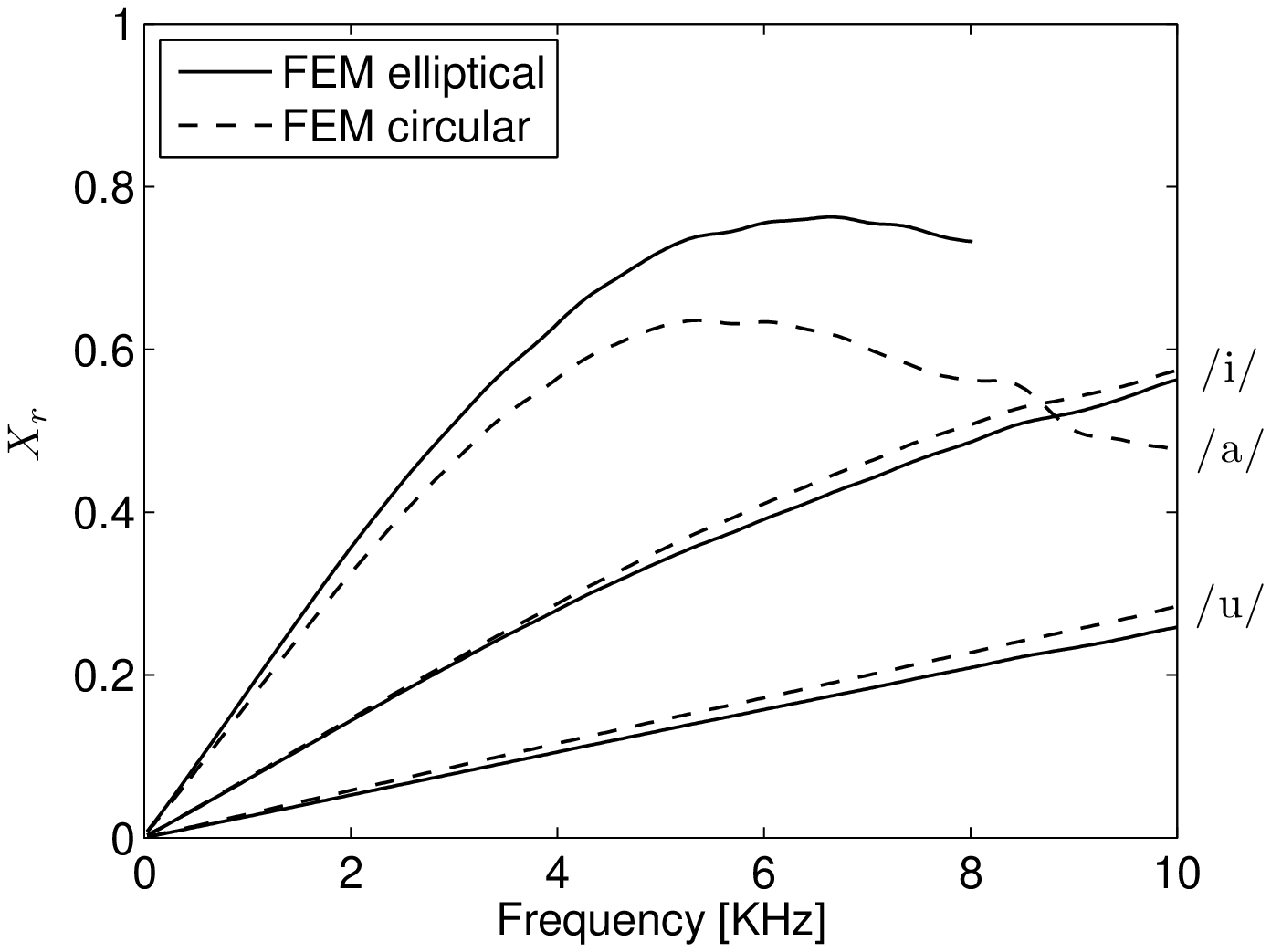}}
  \caption{(a) Radiation resistance and (b) reactance for the elliptical and circular vocal tracts of vowels /a/, /i/ and /u/.}
  \label{Fig:Zr}
\end{figure*}
\c

The above example shows that the condition in Eq.~\eqref{eq:sh} for the microphone spacing is too loose and that stronger requirements are needed in practice for the spacing limiting values. In order to define them we will resort to the so called singularity factor (SF) of the TMTF method, introduced by \cite{Jang98}. The SF indicates the sensitivity of the TMTF method to errors in the input pressures $P_1(f)$ and $P_2(f)$; the higher the SF value the stronger the influence of the error sources in the computed impedances. When computing the SF in the experimental framework, it is assumed that all errors are of the white type, uncorrelated and with constant variance. This will not be the case of a single numerical simulation, the error being totally deterministic i.e., always the same if we repeat the computation. However, if we consider the simulation of a given vocal tract impedance being representative of a certain ensemble average of vocal tracts having e.g., slight different geometry details and material characteristics, we can hypothesize that errors arising from these simulations would satisfy the error requirements for the SF computation. It would then be logical to demand that the chosen spacing for the virtual microphones results in a small SF value.

%
\c
In Fig.~\ref{Fig:Zr3DSF}, we have plotted the SF curve (thick line) for the standard TMTF method in a lossless impedance duct (implementation of the SF for a lossy impedance duct is out of the scope of this work), according to the procedure described in \cite{Jang98}. However, instead of representing the SF dependence with frequency, we have plotted it against the microphone spacing $s$ (normalized by $\lambda_{min}$). Moreover, we have also plotted the microphone spacings corresponding to the resistance and reactance curves in  Fig.~\ref{Fig:Zr3Ds} as vertical lines. As observed in the figure, the SF values for the extreme spacings corresponding to the mesh size and singularity values are much higher than the threshold value of ${\rm SF} \leq 1.7$, recommended in the experimental framework \citep[see][]{Jang98}. According to this criterion we can get a more restrictive range for the virtual microphone spacing than that provided by Eq.~\eqref{eq:sh}, namely
\c
\begin{equation}
h'<0.1 \lambda_{min} <s<0.4\lambda_{min}.
\label{eq:sh'}
\end{equation}
\c
The optimum, and thus recommended, spacing $s$ will be that minimizing SF which is close to $s \simeq 0.25 \lambda_{min}$ in Fig.~\ref{Fig:Zr3DSF}. In our computations, we have used $s=1$~cm (see section~\ref{ss:sim}) that yields $s \simeq0.29 \lambda_{min}$.

\section{Vocal tract impedances} \label{s:IV}
\c
In this section, radiation and input impedances for vowel vocal tracts using the described FEM-TMTF approach will be presented. As detailed in section~\ref{s:II}, we have computed the radiation impedance for vowels /a/, /i/ and /u/, considering circular and elliptical mouth apertures, and the input impedance of vowel /a/ for circular and elliptical vocal tracts.

In Fig.~\ref{Fig:Zr} we present the results for the radiation impedance, split in terms of the radiation resistance (Fig.~\ref{Fig:Rvowels}) and reactance (Fig.~\ref{Fig:Xvowels}).
\c
All results are provided up to $10$~KHz except for the elliptical /a/, the analysis only being valid in this case up to $\sim8$~KHz because of the limiting even mode (2,1) ($f=8.6$~KHz, see Table~\ref{tab:modes}). Note however, that we can reach this value thanks to proper location of the virtual microphones (see section~\ref{ss:planewave}) and that the experimental TMTF would have only let us to measure impedance values up to $f=4.2$~KHz, corresponding to the even mode (0,1) (see Table~\ref{tab:modes}). Similarly, the plane wave frequency restrictions for the circular /a/ ($f=8.21$~KHz) and for the elliptical /i/ ($f=8.27$~KHz) in the experimental TMTF, can be easily overcome thanks to centerline microphone positioning. Thus, for all analyzed cases, except for the elliptical /a/, the radiation impedance can be computed for the whole frequency range of interest $(0\ \ 10]$~KHz, without problems (see Table~\ref{tab:modes}).

The differences between the resistance and reactance curves for the circular and elliptical mouth apertures in Fig.~\ref{Fig:Zr} can be justified as follows (remember that for a given vowel only the shape of the mouth, elliptical or circular, changes but not the total mouth surface). In the plane wave propagation regime, the curves should be almost identical. However, as explained in section~\ref{ss:sim}, the reference surface for elliptical mouth apertures is defined at the intersection between the major semi-axis of the impedance duct with the sphere. This implies that there is small portion of the duct beyond the reference surface, which reaches the intersection between the minor semi-axis and the sphere, and plays somehow the role of some kind of lips. Its effects on the radiation impedance of the elliptical ducts can be understood as those of a modified end correction.
%
\c
To first order approximation, when $R_r^2+X_r^2<<1$, the inertial end correction only affects the reactance. This is what can be observed for vowel /u/, which shows the same resistance values for the circular and elliptical cases (see Fig.~\ref{Fig:Rvowels}), but different ones for the reactance (see Fig.~\ref{Fig:Xvowels}). For vowels /a/ and /i/ the condition $R_r^2+X_r^2<<1$ is no longer satisfied, say for frequencies bigger than $4$~KHz, and differences in resistance values can become more clearly appreciated. It is to be noted that above $4$~KHz, differences between the elliptical and circular cases are not only due to the end correction effects but also to the presence of higher order modes in the impedance ducts.
%

\begin{figure}[!htb]
  \centering
  \includegraphics[height=54mm]{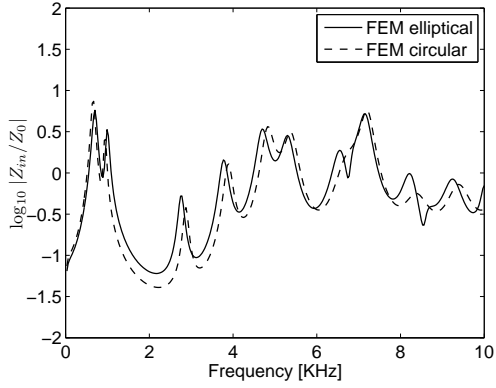}
  \caption{Input impedance for the circular and elliptical~/a/.}
  \label{Fig:Zina}
\end{figure}

Finally, we have computed the input impedance of the circular and elliptical /a/ and plot their moduli in Fig.~\ref{Fig:Zina}.
\c
\c
In contrast to the radiation impedance computation, the input impedance for elliptical /a/ can be computed with no problem up to $10$~KHz. This is so because the glottal section is much smaller than the mouth aperture; a narrower impedance duct is then required, which will have the first non-planar mode beyond $10$~KHz.
\c
Looking at Fig.~\ref{Fig:Zina}, we observe that below $5$~KHz the input impedances moduli of the circular and elliptical /a/ are very similar indicating that we are in the plane wave propagation range. Note however, that there is a certain formant shift to lower frequencies for the elliptical /a/, because its radiation reactance is higher than that of the circular /a/ (see Fig.~\ref{Fig:Xvowels}). 
Above $5$~KHz, differences become notorious due to non-planar high order mode effects.

\section{Conclusions} \label{s:V}
\c
In this paper, we have shown how the experimental two-microphone transfer function method (TMTF) can be adapted to compute impedances of elliptical vocal tracts, from time domain simulations of the irreducible wave equation. This avoids having to compute the acoustic velocity field to do so.


However, using the TMTF in the numerical context it is not straightforward and several considerations have been analyzed. These have resulted in the following observations. First, it is important to impose losses at the impedance duct walls to attenuate the first duct eigenmode and achieve reasonable computational times. This implies using complex wavenumbers in the TMTF expressions, which have been derived for three-dimensional cylindrical impedance ducts with elliptical cross section. Second, we have seen that the frequency range of validity of the experimental TMTF method can be almost doubled by locating the virtual microphones at the impedance duct centerline. This allows overcoming to some extent the plane wave limitation of the experimental TMTF. Third, a range of possible values for the virtual microphone spacing has been proposed making use of the so called singularity factor. An optimum value of a quarter wavelength of the maximum frequency to be solved is recommended.

The radiation impedance of vowels /a/, /i/ and /u/ with circular and elliptical mouth apertures, and the input impedance of vowel /a/, again for the elliptical and circular cases have been computed to test the proposed approach. Simple geometries of vocal tracts available in the literature have been used though any realistic one (e.g., generated from MRI) could have been used instead. All simulations have been carried out using a finite element (FEM) approach with a perfectly matched layer (PML) to allow outward waves from the mouth propagate to infinity.


\c
\begin{acknowledgments}
\c
The authors greatly acknowledge Dr. Ramon Codina for providing us with an initial FEM code for the scalar convection-difusion-reaction equation from which the actual one has evolved and Dr. Francesc Alías for fruitful discussions and information on several topics related to speech synthesis.
\c
The first author would also like to acknowledge the Generalitat de Catalunya (AGAUR) for the pre-doctoral FI Grant No.~${\rm 2012FI\_B~00727}$.
\end{acknowledgments}
\c

\appendix
\c
\section{Time domain FEM approach}\label{AppendixA}
\c
In this appendix we include the details of the time domain FEM approach used to solve problem~\eqref{eq:WaveEq} in section~\ref{s:I}. This consists in extending the finite difference PML formulation for the irreducible wave equation in \cite{Grote10}, to the FEM framework.

If we replace the Sommerfeld radiation condition with the PML, the original acoustic wave equation \eqref{eq:WaveEq} becomes modified to
\begin{subequations}
\begin{align}
  &\partial^2_{tt} p - c^2_0\nabla^2 \ p = \nabla \cdot \phib
  - \alpha \partial_t p
  - \beta p
  - \gamma \psi
 \label{eq:PMLa} \\
  &\partial_{t} \phi_i = -\xi_i \phi_i + c_0^2 a_i \partial_{i} p + c_0^2 b_i \partial_{i} \psi,
  ~ \forall i=1,2,3
  \label{eq:PMLb} \\
  &\partial_t \psi = p
  ,
 \label{eq:PMLc}
\end{align}
in $\Omega,~t>0$, with boundary and initial conditions
\begin{align}
 \nabla p \cdot \n &= -\rho_0/S \partial_t Q \equiv g &{\rm on}& ~ \GammaG, ~t>0, \label{eq:PMLBC1} \\
 \nabla p \cdot \n &= - \mu / c_0 \partial_t p        &{\rm on}& ~ \GammaW, ~t>0, \label{eq:PMLBC2} \\
 \nabla p \cdot \n &= 0                               &{\rm on}& ~ \GammaH, ~t>0, \label{eq:PMLBC3} \\
 \nabla p \cdot \n &= 0                               &{\rm on}& ~ \GammaI, ~t>0, \label{eq:PMLBC4} \\
 p=0     , ~ \partial_t p   &=0                       &{\rm in}&~\Omega, t=0, \label{eq:PMLIC1} \\
 \phi_i=0 , ~ \partial_t \phi_i&=0, ~\forall i=1,2,3  &{\rm in}&~\Omega, t=0, \label{eq:PMLIC2} \\
 \psi =0  , ~ \partial_t \psi &=0                     &{\rm in}&~\Omega, t=0. \label{eq:PMLIC3}
\end{align}
\label{eq:PML}
\end{subequations}
\c
Note that \eqref{eq:PMLa} modifies the right hand side (r.h.s.) of \eqref{eq:WaveEqa} with the inclusion of some extra terms involving the auxiliary functions $\psi$ and $\phib=(\phi_1, \phi_2,\phi_3)$. Moreover, four additional scalar equations \eqref{eq:PMLb}-\eqref{eq:PMLc} are needed for these additional functions.
In what concerns the coefficients $\alpha$, $\beta$, $\gamma$, $a_i$, $b_i$ in \eqref{eq:PML}, they  depend on the damping profiles $\xi_i$ and are given by $\alpha=\xi_1 + \xi_2 + \xi_3$, $\beta=\xi_1 \xi_2 + \xi_2 \xi_3 + \xi_3 \xi_1$, $\gamma=\xi_1 \xi_2 \xi_3$, $a_1=\xi_2+\xi_3-\xi_1$, $a_2=\xi_3+\xi_1-\xi_2$, $a_3=\xi_1+\xi_2-\xi_3$, $b_1=\xi_2\xi_3$, $b_2=\xi_3\xi_1$ and $b_3=\xi_1\xi_2$.
\c
The damping profiles $\xi_i$ are used to control the amount of absorption in the PML and many options exist for them. Following \cite{Grote10}, we have used
\c
\begin{equation}
\xi_i(x_i)=
\hat{\xi}_i \[
\frac{|x_i-l_i|}{L_i}-\frac{{\rm sin}\( \frac{2 \pi |x_i-l_i|}{L_i} \)}{2\pi}
\]
\end{equation}
\c
for $l_i \leq |x_i| \leq l_i+L_i$. $\hat{\xi}_i$ is a constant accounting for the damping effect in the $i$-th direction, $l_i$ is the $i$-th coordinate  of the PML layer and $L_i$ the thickness of the PML region in the $i$-th direction.
\c
The constant $\hat{\xi}_i$ depends on the discretization and thickness of the layer and can be computed as
\c
\begin{equation}
\hat{\xi}_i=\frac{c_0}{L_i} {\rm log} \( \frac{1}{r_{\infty}}\),
\label{eq:PMLhatxi}
\end{equation}
\c
with $r_\infty$ standing for the relative reflection at the boundary of the PML. The PML boundary can be truncated using either a Dirichlet or a Neumann homogeneous condition (the latter has been our choice, see \eqref{eq:PMLBC4}). On the other hand, notice that for $l_i \leq |x_i|$, $\xi_i(x_i)=0$, i.e., outside the PML, the damping profiles $\xi_i$ vanish. In this case, the modified PML wave equation \eqref{eq:PMLa} reduces to the standard wave equation in \eqref{eq:WaveEqa}.

The set of partial differential equations \eqref{eq:PMLa}-\eqref{eq:PMLc} supplemented with boundary and initial conditions \eqref{eq:PMLBC1}-\eqref{eq:PMLIC3} constitute the problem we aim at solving. If a FEM approach is to be used to find a numerical solution for \eqref{eq:PML}, we first have to set this problem in its weak or variational form.  As usual, we first multiply \eqref{eq:PML} by test functions $q,v_i,w$ ($q$ for the pressure, $v_i$ for the first auxiliary functions $\phi_i$, and $w$ for the second auxiliary function $\psi$) and we integrate over the computational domain $\Omega$. Applying the divergence theorem and making use of boundary conditions, we get the weak form of the problem, which consists in finding $p,\phi_i$ and $\psi$ such that
\c
\begin{subequations}
\begin{align}
  \(q,\partial^2_{tt}p \)
  	& + c_0 \(q, \mu \partial_t p\)_{\GammaW} + c^2_0 \(\nabla q,\nabla p\) =
  	c^2_0 \(q, g \)_{\GammaG} \nonumber \\
  	&+ \sum _{i=1}^{3} \(q, \partial_i \phi_i \)
    - \(q,\alpha \partial_t p\) \nonumber \\
  	&- \(q,\beta p\)
  	- \(q,\gamma \psi\)
  	,	\\
  \(v_i,\partial_t \phi_i\)&= - \(v_i,\xi_i \phi_i\) + c_0^2 \(v_i,a_i \partial_i p\) \nonumber \\
  	&+ c_0^2 (v_i,b_i \partial_i \psi)
    , ~\forall i=1,2,3, \\
  \(w,\partial_t \psi\) &= \(w,p\),
\end{align}
\c
in $\Omega,~t>0$, with initial conditions
\c
\begin{align}
 \(q,p\)&=0        , ~ \(q,\partial_t p\)      =0,                        \\
 \(v_i,\phi_i\)&=0 , ~ \(v_i,\partial_t \phi_i\)=0,    ~\forall i=1,2,3,   \\
 \(w,\psi\) &=0     , ~ \(w,\partial_t \psi\)  =0,
\end{align}
\label{eq:weak}
\end{subequations}
\c
in $\Omega, t=0$, for all $q,v_i,w$.
\c
To shorten notation, in \eqref{eq:weak} the integral of the product of any two functions $f,g$ in $\Omega$ has been written as
\c
\begin{align}
\( f,g \):=\int_{\Omega}f g d\Omega,
\label{eq:Prod-L2}
\end{align}
\c
whilst integrals over boundaries have been explicitly indicated e.g., $\(f,g\)_{\Gamma_G}$.

To find a numerical solution to problem \eqref{eq:weak}, we have to discretize it both in space and time.  Let us first discretize it in space using a FEM formulation.
\c
Given a finite element partition of $\Omega$ with $n_{el}$ elements and $n_p$ nodes and discretizing problem \eqref{eq:weak} following the Galerkin method, we have expanded the discrete versions of the unknowns $p$, $\phi_{i}$, $\psi$ and the discrete versions of the test functions $q$, $v_{i}$ and $w$ in piecewise linear terms of piecewise linear shape functions $N(\x)$ (e.g., $p_h=\sum_{b=1}^{n_p} N^{b} P^{b}$), so that we get the following time evolving algebraic matrix system
\begin{subequations}
\begin{align}
\M \ddot{\P} &+ c_0 \B \dot{\P} +c^2_0 \K \P = c^2_0 \L \nonumber \\
&+ \sum _{i=1}^{3} \Bi \Phii
- \Ma \dot{\P}
- \Mb \P
- \Mg \Psib,
\\
\M \Phiidot &=-\Mxii \Phii + c^2_0 \Biai \P \nonumber \\
&+ c^2_0 \Bibi \Psib, ~\forall i=1,2,3,
\\
\dot{\Psib} &=\P,
\end{align}
\c
in $\Omega,~t>0$, with initial conditions
\c
\begin{align}
 \P &=0       , ~ \dot{\P}   =0, \\
 \Phii  &=0   , ~ \Phiidot   =0,    ~\forall i=1,2,3,   \\
 \Psib  &=0   , ~ \dot{\Psib} =0,
\end{align}
\c
in $\Omega, t=0$.
\c
\label{eq:GarlekinMat}
\end{subequations}
\c
In \eqref{eq:GarlekinMat}, $\P$, $\Phii$ and $\Psib$ stand for the vectors of nodal values that respectively correspond to the pressure and auxiliary functions  (e.g., $\P=(P^1 \cdots P^{n_p})^\top$), whereas the remaining matrix and vector entries are given by
\begin{subequations}
\begin {align}
M^{ab} &=\(N^a,N^b\),                             &M_\alpha^{ab} &=\(N^a,\alpha_h N^b\),     \label{eq:MatrixA}
\\
M_\beta^{ab} &= \(N^a,\beta_h N^b\),              &M_\gamma^{ab} &= \(N^a,\gamma_h N^b\),    \label{eq:MatrixB}
\\
M_{\xi_i}^{ab} &=\(N^a,\xi_{ih} N^b\),           &B^{ab} &=\(N^a, \mu N^b\)_{\GammaW},\label{eq:MatrixC}
\\
B_{i,a_i}^{ab} &=\(N^a,a_{ih} \partial_i N^b\),  &B_{i}^{ab} &=\(N^a,\partial_i N^b\),       \label{eq:MatrixD}
\\
B_{i,b_i}^{ab} &=\(N^a,b_{ih} \partial_i N^b\),  &K^{ab} &=\(\nabla N^a, \nabla N^b \),      \label{eq:MatrixE}
\\
L^{a} &=\(N^a, g \)_{\GammaG}.
\end {align}
\label{eq:matrix}
\end{subequations}
As usual, the domain integrals in \eqref{eq:matrix} are to be understood as the summation of integrals over elements $\Omegae$, i.e., $(\cdot,\cdot)_{\Omega}=\sum_{e=1}^{n_{el}} (\cdot,\cdot)_{\Omegae}$. The subscripts $h$ denote the discrete versions of the corresponding continuous variables.

Let us next proceed to the time discretization of \eqref{eq:GarlekinMat}. A finite difference approach has been used to do so. We consider a constant time step ($\Delta t = t^{n+1}-t^n$) discretization of the time interval $(0,T)$ into $0<t^1<\cdots<t^{n-1}<t^{n}<t^{n+1}<\cdots<t^N\equiv T$. A second order finite difference central scheme has then been implemented for the pressure time derivatives, whilst a first order central scheme has been used for the time derivatives of the auxiliary variables in the PML region. This is so because first order schemes are known to introduce strong numerical dissipation, which in this case is advantageous to help absorbing the waves crossing the PML.

Inserting these time derivative approximations in \eqref{eq:GarlekinMat} yields
\begin{subequations}
\begin{align}
  \M &\frac{\P^{n+1} -2 \P^n + \P^{n-1}}{\Delta t^2}
  + c_0 \B \frac{\P^{n+1}-\P^{n-1}}{2 \Delta t}
  \nonumber \\
  &+c^2_0 \K \P^n = c^2_0 \L^n
  + \sum _{i=1}^{3} \Bi \Phii^n
  - \Ma \frac{\P^{n+1}-\P^{n-1}}{2 \Delta t}
  \nonumber \\
  &- \Mb \P^n
  - \Mg \frac{\Psib^{n+1/2}+\Psib^{n-1/2}}{2},
\label{eq:TimeScheme1a}
\end{align}
\begin{align}
  \M &\frac{\Phii^{n+1}-\Phii^{n}}{\Delta t} =-\Mxii \frac{\Phii^{n+1}+\Phii^{n}}{2}
  \nonumber \\
  &+ c^2_0 \Biai \frac{\P^{n+1}+\P^{n}}{2}
  + c^2_0 \Bibi \Psib^{n+1/2},
\label{eq:TimeScheme1b}
\end{align}
\begin{align}
  \frac{\Psib^{n+1/2}-\Psib^{n-1/2}}{\Delta t} &=\P^n,
\label{eq:TimeScheme1c}
\end{align}
with initial conditions
\begin{align}
 \P^0 &=0     , ~ \P^1    =0, \\
 \Phii^0 &=0   , ~\Phii^1 =0,    ~\forall i=1,2,3,   \\
 \Psib^0 &=0   , ~\Psib^1 =0.
\end{align}
\label{eq:TimeScheme1}
\end{subequations}
Note that \eqref{eq:TimeScheme1} corresponds to a purely explicit scheme, where all the unknowns can be calculated at time step $n+1$ from values already known from previous steps. Solving \eqref{eq:TimeScheme1} at time step $t=n+1$ involves
\c
\begin{enumerate}
	\item Use \eqref{eq:TimeScheme1c} to compute $\Psib^{n+1/2}$
	\item Insert $\Psib^{n+1/2}$ into \eqref{eq:TimeScheme1a} and compute $\P^{n+1}$
	\item Update $\Phii^{n+1}$ using \eqref{eq:TimeScheme1b},  	
	      $\Psib^{n+1/2}$ and $\P^{n+1}$
\end{enumerate}
\c
To fasten all computations, matrix inversion is avoided as usual by means of a lumped approximation for all mass matrices \eqref{eq:MatrixA}-\eqref{eq:MatrixC} \citep{HughesBook}.

The system of equations \eqref{eq:TimeScheme1} constitutes the final numerical scheme that has been used for all the computational simulations in this work.

\end{document}